\documentclass[aps,prl,twocolumn,10pt]{revtex4-2}

\usepackage{amssymb}
\usepackage{graphicx}
\usepackage{amsmath}
\usepackage[bookmarks = false,colorlinks = true,allcolors = blue]{hyperref}
\usepackage{color,xcolor}
\usepackage{multirow}
\usepackage{ulem}
\usepackage{braket}
\usepackage{verbatim}
\usepackage{array}
\usepackage{upgreek}

\begin{document}

\title{Deterministic Time-Bin Entanglement between a Single Photon and an Atomic Ensemble}

\author{Peng-Fei~Sun$^{1,\,2}$}
\author{Yong Yu$^{1,\,2}$}
\author{Zi-Ye~An$^{1,\,2}$}
\author{Jun~Li$^{1,\,2}$}
\author{Chao-Wei~Yang$^{1,\,2}$}
\author{Xiao-Hui Bao$^{1,\,2}$}
\author{Jian-Wei Pan$^{1,\,2}$}

\affiliation{$^1$Hefei National Laboratory for Physical Sciences at Microscale and Department
of Modern Physics, University of Science and Technology of China, Hefei,
Anhui 230026, China}
\affiliation{$^2$CAS Center for Excellence in Quantum Information and Quantum Physics, University of Science and Technology of China, Hefei, Anhui 230026, China}

\begin{abstract}
Hybrid matter-photon entanglement is the building block for quantum networks. It is very favorable if the entanglement can be prepared with a high probability. In this paper, we report the deterministic creation of entanglement between an atomic ensemble and a single photon by harnessing Rydberg blockade. We design a scheme that creates entanglement between a single photon's temporal modes and the Rydberg levels that host a collective excitation, using a process of cyclical retrieving and patching. The hybrid entanglement is tested via retrieving the atomic excitation as a second photon and performing correlation measurements, which suggest an entanglement fidelity of 87.8\%. Our source of matter-photon entanglement will enable the entangling of remote quantum memories with much higher efficiency.
\end{abstract}

\maketitle

The construction of quantum networks~\cite{Kimble2008,Simon2017,Wehner2018}
over long distance requires matter qubits with an efficient optical interface, since photons are an excellent candidate for long-distance transmission, and matter qubits are necessary to connect different segments. A paradigm for this situation is an entanglement pair between a single photon and a matter qubit. For application in quantum networks, it is necessary that the hybrid entanglement has a very high preparation efficiency and fidelity, at the same time the matter qubit can be preserved for a long duration~\cite{Sangouard2011}.

It is very promising of harnessing atomic ensembles~\cite{Duan2001,Sangouard2011,Hammerer2010} for quantum networks, since an ensemble of atoms are relatively easier to prepare in comparison with single atoms, and it has additional advantage of collective enhancement, which enables efficient interaction with single photons. In the past years, plenty of efforts have been paid to improving the 
performances of laser-cooled atomic ensembles~\cite{Simon2007,Zhao2009,zhao2009long,Lan2009,Dudin2010,Radnaev2010,Bao2012,Dudin2013,xu_long_2013,cho_highly_2016,Yang2016,Pu2017,tian_spatial_2017,chrapkiewicz_high-capacity_2017,Shi2018,Hsiao2018,vernaz-gris2018,Wang2019,chang_long-distance_2019,jing_entanglement_2019,cao_efficient_2020,yu_entanglement_2020,wang_cavity-enhanced_2021}. Notably, very high retrieval efficiency has been achieved either by using large optical depth~\cite{cho_highly_2016,Hsiao2018,Wang2019,cao_efficient_2020} or with the assistance of a low-finesse cavity~\cite{Simon2007,Bao2012,Yang2016,wang_cavity-enhanced_2021}, sub-second storage has been achieved by confining the atoms with optical lattice and compensating the differential light shift~\cite{Dudin2010,Radnaev2010,Yang2016,wang_cavity-enhanced_2021}, telecom frequency conversion has also been employed to extend distance of remote entanglement~\cite{yu_entanglement_2020}.

When atomic ensembles are employed as quantum memories, they are typically modeled as non-interacting atoms, and collision between atoms are unfavorable. In this situation, the preparation of entanglement with a single photon is probabilistic, and the preparation probability is kept very low typically to minimize the contribution of high-order events~\cite{Duan2001,Sangouard2011}. However, Rydberg interaction provides a promising solution to create entanglement deterministically~\cite{Zhao2010c,Han2010}, as a single atom excited in a Rydberg state can block further excitations into the Rydberg state~\cite{Lukin2001,Saffman2010c}. In recent years, plenty of research have been performed on Rydberg quantum optics~\cite{Peyronel2012a,Dudin2012e,Dudin2012sc,Firstenberg2013a,Li2013,tiarks_single-photon_2014,gorniaczyk_single-photon_2014,Ebert2014,ebert_coherence_2015,Li2016,gorniaczyk_enhancement_2016,thompson_symmetry-protected_2017,busche_contactless_2017,tiarks_photonphoton_2019,Li2019,cantu_repulsive_2020}. In particular, L. Li \textit{et al}~\cite{Li2013} demonstrated entanglement between a Rydberg excitation and a number-state encoded light field. The number-state encoding suffers from being phase sensitive. To avoid this, J. Li \textit{et al} realized entanglement between a Rydberg excitation and a polarization encoded single photon~\cite{Li2019}. This scheme nevertheless succeeds merely 50\% in principle, limiting its efficiency in entangling remote nodes~\cite{yuan2008experimental,yu_entanglement_2020} via entanglement swapping~\cite{Pan1998}.

\begin{figure*}[htb]
    \centering
    \includegraphics[width=\textwidth]{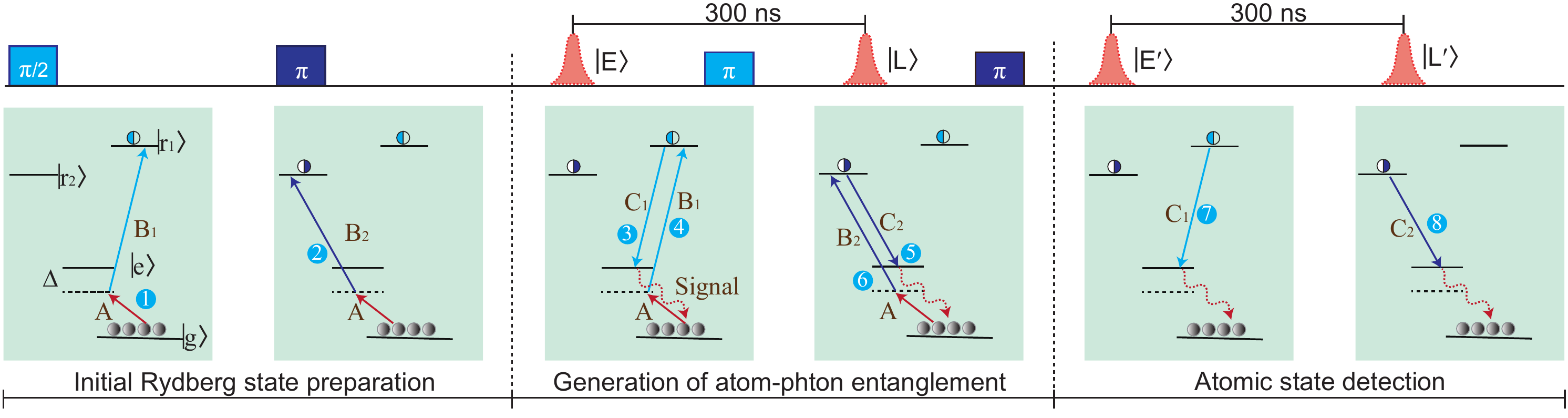}
    \caption{Scheme of generating deterministic time-bin entanglement. In the first phase, we prepare a Rydberg superposition state $(|R_1\rangle+|R_2\rangle)/\sqrt{2}$. In the second phase, we apply the cyclical ``retrieving and patching'' process for $|R_1\rangle$ and $|R_2\rangle$ in sequence, which results in retrieval photon whose temporal mode is entangled with the Rydberg levels of the collective excitation. In the last phase, we detect the atomic state via retrieving it to the temporal mode of a second photon. The excitation and patching process involve a two-photon Raman process. Spatial arrangement of the Raman beams is shown in Fig.~\ref{setup}. Numerically filled circles indicate relative time sequences of all the steps.}\label{scheme}
\end{figure*}

In this paper, we propose and experimentally realize a scheme that entangles an atomic ensemble with a single photon's time-bin modes deterministically. The detailed scheme is depicted in Fig.~\ref{scheme}. We consider a mesoscopic atomic ensemble in the regime of full blockade~\cite{Saffman2010c}. All atoms initially stay in a ground state $|g\rangle$. Two Rydberg levels $|r_1\rangle$ and $|r_2\rangle$ are employed to host a collective excitation. Our scheme starts from a Rydberg state preparation phase, in which we first apply a collective $\pi/2$ pulse coupling $|g\rangle$ and $|r_1\rangle$ resulting in a collective state of $(|R_1\rangle + |G\rangle)/\sqrt{2}$, and apply a subsequent $\pi$ pulse coupling $|g\rangle$ and $|r_2\rangle$ resulting in a superposition state of $(|R_1\rangle + |R_2\rangle)/\sqrt{2}$, where $|G\rangle$ denotes all atoms in $|g\rangle$ and $|R_{j}\rangle = N^{-1/2}\sum_{i=1}^{N}|g^{1}g^{2}\cdots r^i_{j}\cdots g^{N}\rangle$ denotes a collective excitation in $|r_j\rangle$, with $N$ being number of atoms and $i$ being an atom index. We name the 2nd $\pi$ pulse as a ``patching'' pulse, since it creates a full excitation out of a half excitation, harnessing Rydberg blockade between $|r_1\rangle$ and $|r_2\rangle$. In the next phase of our scheme, atom-photon entanglement is generated via a cyclical process of ``retrieving and patching''. We first retrieve $|R_1\rangle$ to a single photon and patching it back by applying a $\pi$ pulse coupling $|g\rangle$ and $|r_1\rangle$. Rydberg blockade between $|r_1\rangle$ and $|r_2\rangle$ again plays a key role, ensuring that a $|R_1\rangle$ excitation is recreated only if there was a $|R_1\rangle$ excitation previously. Afterwards, we apply the retrieving and patching process for $|R_2\rangle$, creating a photon in a delayed temporal mode. After these steps, the joint atom-photon state can be expressed as
\begin{equation}
    |\Psi_{ap}\rangle=(1/\sqrt{2})(|R_{1}\rangle|E\rangle+|R_2\rangle|L\rangle),
    \label{eq:1A}
\end{equation}
where $|E\rangle$ or $|L\rangle$ denotes a single photon in the temporal early or late mode. We note that all the atomic manipulation steps are deterministic. The retrieving process from a Rydberg excitation to a single photon~\cite{Saffman2002} is also deterministic in principle, as long as the optical depth is high enough. Thus we claim that our scheme is deterministic, which differentiates from the traditional DLCZ scheme~\cite{Duan2001} that creates entanglement probabilistically through spontaneous Raman scattering. In the last phase, we retrieve the atomic state $|R_1\rangle$ and $|R_2\rangle$ in sequence to a second photon in the temporal mode of $|E'\rangle$ and $|L'\rangle$ to verify the entanglement. 

\begin{figure}[htbp]
\centering
\includegraphics[width=1\columnwidth]{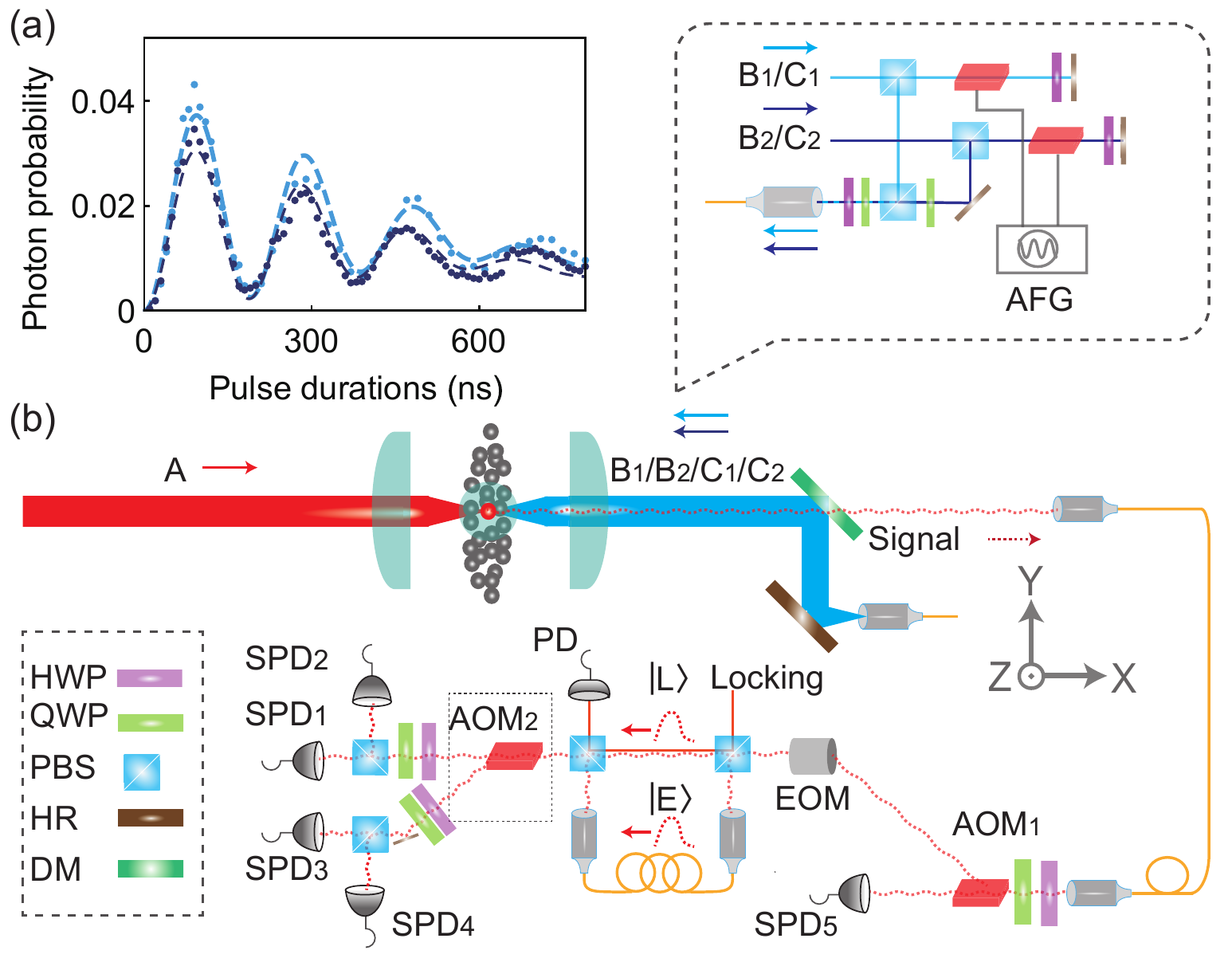}
\caption{Experimental setup. (a) Observed collective Rabi oscillations. Light blue dots represent the oscillation between $|G\rangle$ and $|R_{1}\rangle$ and dark blue dots represent the oscillation between $|G\rangle$ and $|R_{2}\rangle$. Dashed lines are the fitting results, indicating a $\pi$ pulse duration of 92.95 ns and 92.18 ns for $|R_{1}\rangle$ and $|R_{2}\rangle$ respectively. (b) Layout of the experiment. The 795~nm beam counterpropagates with the 474~nm beams, all of which are tightly focused to address a small region of the atomic cloud to define a mesoscopic ensemble with full blockade (more details available in \cite{Zhou2020}). Single photons are retrieved in the $x$ direction, afterwards subjected to fiber coupling, temporal filtering, interfering through an unbalanced MZ interferometer, switching, rotating in polarization and final detecting sequencially. The fibers used in the two arms of the the MZ interferometer are 65 m and 5 m which give a relative photon delay of about 300 ns. The interferometer is actively stabilized  by injecting a locking beam into the idle port. Our experiment involves four blue beams ($\mathrm{B_{1}/B_{2}/C_{1}/C_{2}}$), which differentiate each other in polarization
, frequency, or both. We make use of a multiplexing scheme shown as inset of Fig.~\ref{setup}b to adjust the frequencies dynamically and combine the beams into a single-mode fiber. AOM$_2$ is merely used for the Bell measurement. Some abbreviations: halfwave plate (HWP), quarterwave plate (QWP), polarizing beamsplitter (PBS), high reflection mirror (HR), dichroic mirror (DM), photodiode (PD), electro-optic modulator (EOM).}\label{setup}
\end{figure}

Our experimental setup is illustrated in Fig.~\ref{setup}. An ensemble of $^{87}$Rb atoms is captured by a conventional magneto-optical trap~(MOT)  lasting 30 ms, and a following dark spontaneous force optical trap~(dark SPOT)~\cite{Ketterle1993,Townsend1996} lasting 70 ms. Afterwards, the atoms are loaded into an optical dipole trap formed by a 1064 nm laser propagating along the $y$ direction with a power of 2 W and a waist of 5 $\upmu$m in the $x$ direction and 25 $\upmu$m in the $z$ direction, resulting in an optical depth of 1.7 along the $x$ direction and a density of $10^{11}\,\mathrm{cm^{-3}}$. The energy levels include a ground state $|g\rangle=|5S_{1/2},\;F=2,\;m_{F}=+2\rangle$, an intermediate state $|e\rangle=|5P_{1/2}, \;F=1,\;m_{F}=+1\rangle$, and two Rydberg states $|r_{1}\rangle=|81S_{1/2},\;m_{j}=+1/2\rangle$ and $|r_{2}\rangle=|81S_{1/2},\;m_{j}=-1/2\rangle$. We employ a two-photon Raman process to excite an atom to a Rydberg level, involving a 795~nm laser and a 474~nm laser. Single-photon detuning is set to $\Delta = -40$ MHz. The two Rydberg levels are selected to be close to each other. Thus we are able to share the same 474 nm laser instead of seeking a second blue laser for the other Rydberg state. Determined by the phase-matching condition~\cite{Saffman2002}, retrieved single photons propagate in the $x$ direction, which is the same as the 795~nm Raman laser. At the same time, the single photons also have the same polarization as the Raman laser, thus slight leakage of the Raman laser will spoil the single photons. In our experiment, we solve this problem by using a temporal filter involving an acousto-optic modulator (AOM$_1$ in Fig.~\ref{setup}b)) to deflect the optical field only when retrieving the single photons. Afterwards, the single photon enters an active unbalanced Mach–Zehnder~(MZ) interferometer, in which the early mode goes through the long arm while the late mode goes through the short arm. The MZ interferometer is employed not only to measure the first photon but also the second photon retrieved for atomic state detection, which is enabled with an optical switch (AOM$_2$ in Fig.~\ref{setup}b). Finally, the photons are measured with single-photon detectors (SPDs) and analyzed in polarization. 

\begin{figure}[htbp]
\centering
\includegraphics[width=.9\columnwidth]{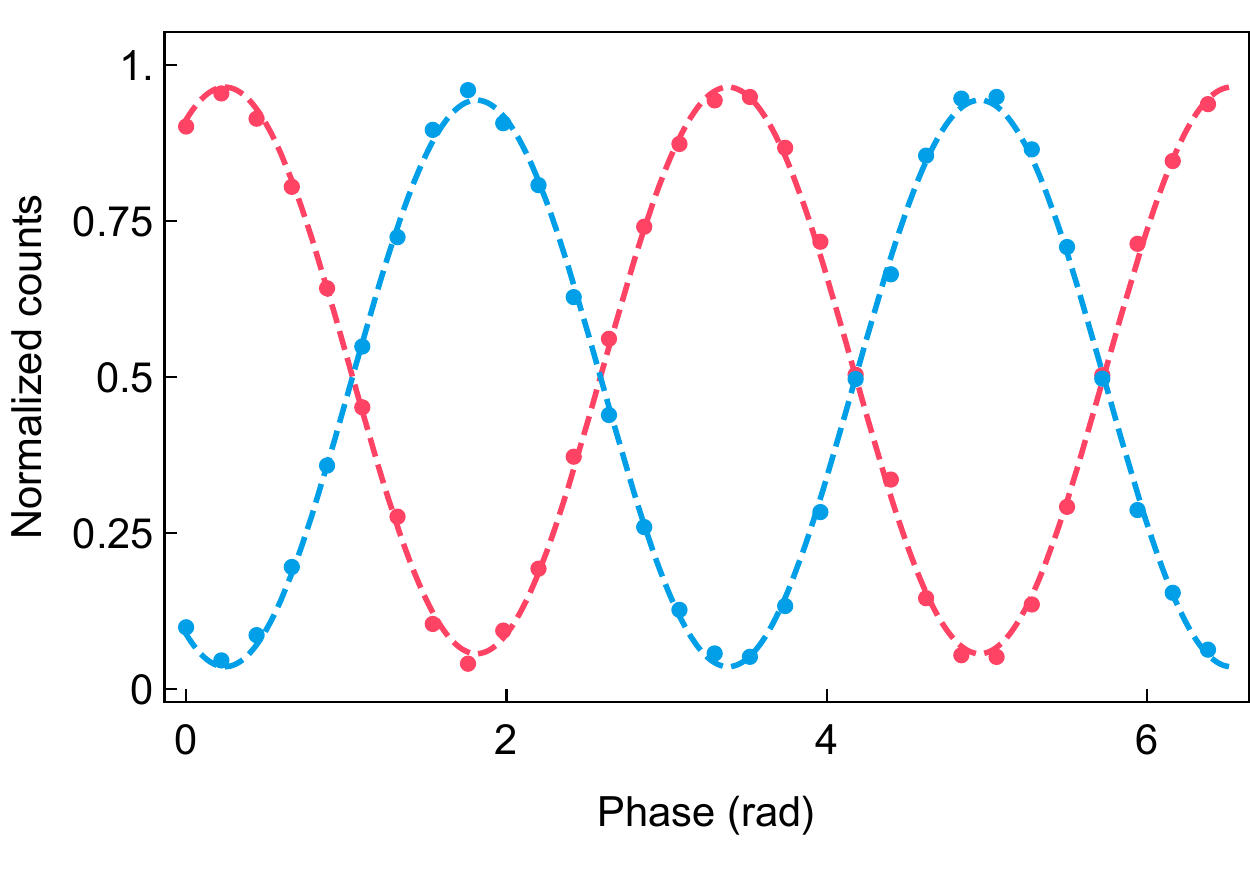}
\caption{Verification of the initial Rydberg state preparation. $(|R_1\rangle+|R_{2}\rangle)/\sqrt{2}$ is converted to $(|E'\rangle+e^{i\phi}|L'\rangle)/\sqrt{2}$ and measured in the basis of $(|E'\rangle \pm |L'\rangle)/\sqrt{2}$. Dependence on the internal phase $\phi$ is plotted. Dashed lines are the fitting results.}\label{interference}
\end{figure}

\begin{figure}[htbp]
    \centering
    \includegraphics[width=.9\columnwidth]{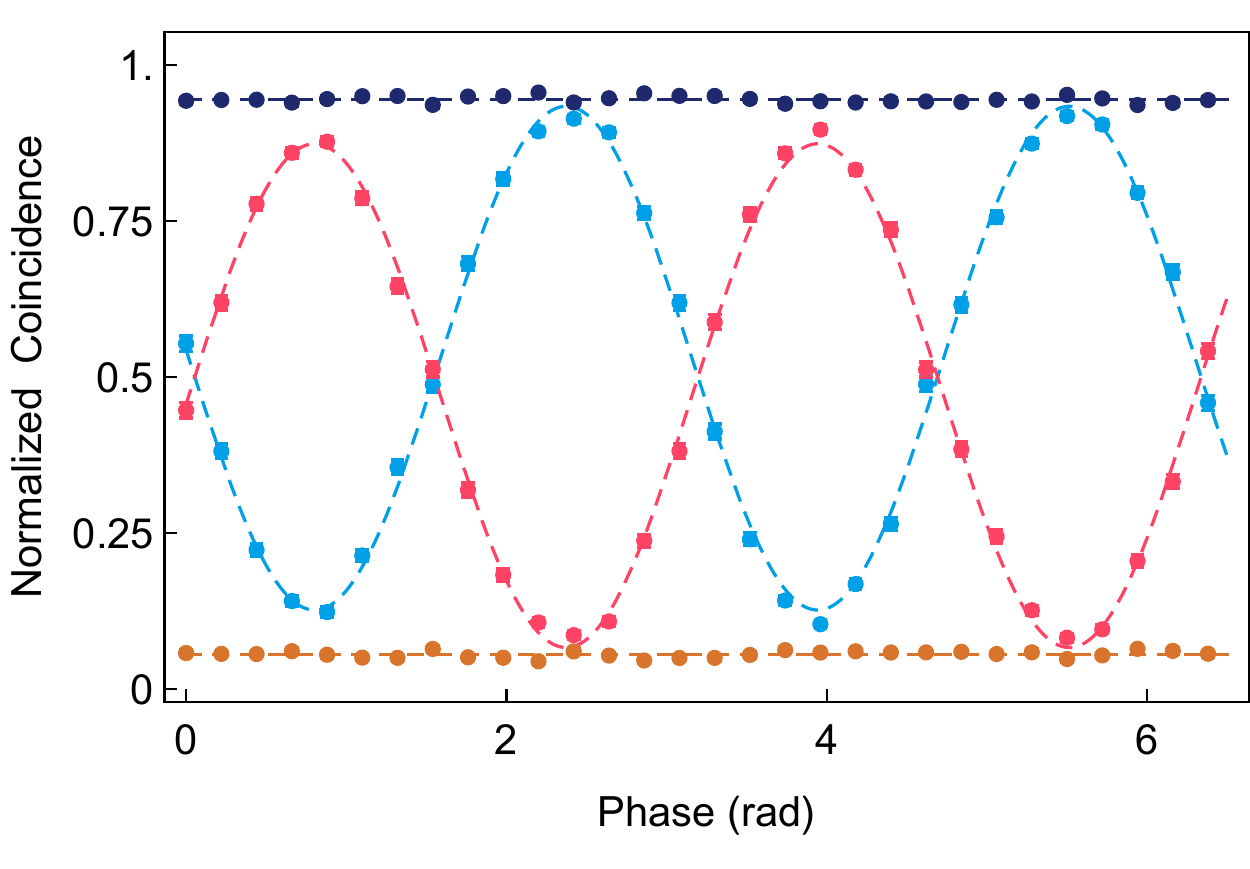}
    \caption{Characterization of the entanglement $|\Psi_{pp}\rangle$ via correlation measurement. In the eigenbasis, the sum of $|E\rangle|E'\rangle$ and $|L\rangle|L'\rangle$ coincidences is shown in dark blue, while sum of $|E\rangle|L'\rangle$ and $|L\rangle|E'\rangle$ coincidences is shown in orange. In the superposition basis, The sum of $|+\rangle|+'\rangle$ and $|-\rangle|-'\rangle$ coincidences is shown in red, while sum of $|+\rangle|-'\rangle$ and $|-\rangle|+'\rangle$ is shown in light blue. The dash lines are the fitting results.}\label{vis} 
\end{figure}

Before characterizing the atom-photon entanglement, it is crucial to verify that the ``patching'' process is coherent and a Rydberg excitation is genuinely created in a superposition state. Therefore, we first perform an experiment that retrieves the atomic excitation immediately after its creation. In other words, we skip the second phase in Fig.~\ref{scheme}. Thus the initial atomic state $(|R_1\rangle+|R_{2}\rangle)/\sqrt{2}$ is directly converted to $(|E'\rangle+e^{i\phi}|L'\rangle)/\sqrt{2}$, with an internal phase $\phi$ being dependent on the relative phases of $\mathrm{B_1}$, $\mathrm{B_2}$, $\mathrm{C_1}$, and $\mathrm{C_2}$. In order to get a stable phase, we make use of an arbitrary function generator (AFG) with dual outputs to control the AOMs for $\mathrm{B_1}/\mathrm{C_1}$ and $\mathrm{B_2}/\mathrm{C_2}$ respectively as shown in the inset of Fig.~\ref{setup}b, which sets a common phase for all the 4 pulses. We perform measurement in the basis of $|\pm\rangle = (|E'\rangle \pm |L'\rangle)/\sqrt{2}$, with the result shown in Fig.~\ref{interference}. We vary the phase of $\mathrm{C_1}$ and measure the single photon counts. As expected, the relative counts show as complementary sinusoidal oscillations as a function of phase. By fitting the result, we get an interference visibility of $V_{+/-}=0.909 \pm 0.006$, which is mainly limited by 
inhomogeneity of the two temporal modes, optical misalignment, and detector aflterpulse. The overall measured photon efficiency is about 1.7\%, which includes preparation efficiency of a Rydberg excitation ($\sim$90\%), retrieval efficiency ($\sim$13\%), transmission and fiber coupling efficiency ($\sim$69\%), AOM deflecting efficiency ($\sim$77\%), transmission efficiency through unbalanced MZ interferometer ($\sim$47\%), and detector efficiency ($\sim$60\%). We note that by harnessing enhancement with a low-finesse ring cavity~\cite{Bao2012}, the retrieval efficiency may be improved significantly. 

Next, we implement the whole scheme shown in Fig.~\ref{scheme}. We add back the phase of atom-photon entanglement generation, by applying the retrieving and patching process for $|R_1\rangle$ and $|R_2\rangle$ in sequence, which creates a first photon that is entangled with atomic excitation in the form of $|\Psi_{ap}\rangle$. To verify the entanglement, we retrieve the atomic excitation as a second photon that entangles with the first photon in the form of $|\Psi_{pp}\rangle=(1/\sqrt{2})(|E\rangle|E'\rangle+e^{i\psi}|L\rangle|L'\rangle)$, with an internal phase $\psi$ dependent on the Raman and retrieving fields. To verify the entanglement, we perform correlation measurement both in the eigenbasis of $|E\rangle/|L\rangle$, and in a superposition basis of $|+\rangle/|-\rangle$. The detectors $\mathrm{SPD_1}$ and $\mathrm{SPD_2}$ are employed for this measurement through temporal multiplexing, and measurement settings are configured by rotating the waveplates. Experimental results are shown in Fig.~\ref{vis}. As expected, when we vary the internal phase $\psi$ linearly, coincidences in the eigenbasis show no dependence, while coincidences in the superposition basis show as sinusoidal oscillations. By fitting the results, we get a visibility of $V_1=0.890 \pm 0.010$ for the eigenbasis, and $V_2=0.811 \pm 0.008$ for the superposition basis, which together give an entanglement fidelity as $F\approx(1+V_1+2V_2)/4=0.878 \pm 0.005$~\cite{Guhne2009}. A moderate visibility in the superposition basis contributes most for the entanglement infidelity. Feasible measures for improvement includes using faster switches, using detectors with a lower probability of afterpulse, and using two Rydberg levels of different principal quantum number to reduce crosstalk. 

\begin{table}[htbp]
    \caption{Measurement results for Bell-CHSH test.}
    \begin{tabular*}{\columnwidth}{@{\extracolsep\fill}ccccc}\toprule
    $\alpha$,\,$\beta$ & 22.5,\,45 & 22.5,\,0 & 67.5,\,45 & 67.5,\,0 \\
    \hline
    $E(\alpha,\beta)$ & -0.415 & -0.531 & -0.568 & 0.659 \\
    \hline
    $\sigma_{E}$ & 0.029 & 0.028 & 0.026 & 0.025 \\
    \toprule\end{tabular*}
    \label{CHSH}
\end{table}

Finally, we verify the entanglement directly via testing the Bell-CHSH inequality~\cite{bell_problem_1966,clauser_proposed_1969}. In this test, the two photons need to be measured in different settings, thus the previous detector multiplexing scheme does not work. We make use of an active switcher (AOM$_2$) to solve this problem. The AOM lets the first photon pass through and directs it to SPD$_1$ and SPD$_2$, while deflects the second photon and directs it to SPD$_3$ and SPD$_4$. To perform the Bell test, we need to measure the $S$ parameter that is defined as $S=|E(\alpha,\beta)+E(\alpha^*,\beta)+E(\alpha,\beta^*)-E(\alpha^*,\beta^*)|$, with $E(\alpha,\beta)$ being the joint expectation value when one photon is measured in the $\alpha$ setting and the other photon is measured in the $\beta$ setting. A setting $\theta$ refers to a measurement in the basis of $|\theta\rangle = \cos(\theta) |E\rangle + \sin(\theta) |L\rangle$ and $|\theta^\perp\rangle = \sin(\theta) |E\rangle - \cos(\theta) |L\rangle$. $E(\alpha,\beta)$ takes a maximal value $1$ for perfect parallel correlation, and $-1$ for perfect anticorrelation. For perfect entanglement and optimized settings, the $S$ parameter can take a maximal value of $2\sqrt{2}$. The threshold to certify entanglement is 
$S>2$. We set the internal phase $\psi$ of $|\Psi_{pp}\rangle$ to $0$ and perform the measurements. The results are shown in Tab.~\ref{CHSH}, and we get $S = 2.173\pm0.055$, which violates the inequality by 3.16 standard deviations. 

In summary, we propose and experimentally realize a scheme of deterministic time-bin entanglement between a single photon and a mesoscopic atomic ensemble. We harness two Rydberg levels to host a collective atomic qubit, and employ a cyclical ``retrieving and patching'' process to create atom-photon entanglement. Self-blockade in each Rydberg level, and cross blockade between the two levels play key roles for the scheme. Our experiment fully implements the scheme. Via performing correlation measurements, we estimate an entanglement fidelity of 87.8\%. We also perform a Bell inequality test to verify the entanglement directly. Further improvements include using a low-finesse cavity to improve the retrieval efficiency~\cite{Simon2007,Bao2012}, transferring the atomic qubit from Rydberg levels to ground state levels~\cite{Ebert2014,Li2016}, and using optical lattice to achieve long-lived storage~\cite{Yang2016,wang_cavity-enhanced_2021} etc. With these developments, the atom-photon entanglement created in our work may become a strong candidate as a fundamental element for constructing large-scale quantum networks~\cite{Kimble2008,Sangouard2011,Simon2017,Wehner2018}. 

This work was supported by National Key R\&D Program of China (No. 2017YFA0303902), Anhui Initiative in Quantum Information Technologies, National Natural Science Foundation of China, and the Chinese Academy of Sciences.

\bibliography{myref}

\begin{thebibliography}{58}%
\makeatletter
\providecommand \@ifxundefined [1]{%
 \@ifx{#1\undefined}
}%
\providecommand \@ifnum [1]{%
 \ifnum #1\expandafter \@firstoftwo
 \else \expandafter \@secondoftwo
 \fi
}%
\providecommand \@ifx [1]{%
 \ifx #1\expandafter \@firstoftwo
 \else \expandafter \@secondoftwo
 \fi
}%
\providecommand \natexlab [1]{#1}%
\providecommand \enquote  [1]{``#1''}%
\providecommand \bibnamefont  [1]{#1}%
\providecommand \bibfnamefont [1]{#1}%
\providecommand \citenamefont [1]{#1}%
\providecommand \href@noop [0]{\@secondoftwo}%
\providecommand \href [0]{\begingroup \@sanitize@url \@href}%
\providecommand \@href[1]{\@@startlink{#1}\@@href}%
\providecommand \@@href[1]{\endgroup#1\@@endlink}%
\providecommand \@sanitize@url [0]{\catcode `\\12\catcode `\$12\catcode
  `\&12\catcode `\#12\catcode `\^12\catcode `\_12\catcode `\%12\relax}%
\providecommand \@@startlink[1]{}%
\providecommand \@@endlink[0]{}%
\providecommand \url  [0]{\begingroup\@sanitize@url \@url }%
\providecommand \@url [1]{\endgroup\@href {#1}{\urlprefix }}%
\providecommand \urlprefix  [0]{URL }%
\providecommand \Eprint [0]{\href }%
\providecommand \doibase [0]{https://doi.org/}%
\providecommand \selectlanguage [0]{\@gobble}%
\providecommand \bibinfo  [0]{\@secondoftwo}%
\providecommand \bibfield  [0]{\@secondoftwo}%
\providecommand \translation [1]{[#1]}%
\providecommand \BibitemOpen [0]{}%
\providecommand \bibitemStop [0]{}%
\providecommand \bibitemNoStop [0]{.\EOS\space}%
\providecommand \EOS [0]{\spacefactor3000\relax}%
\providecommand \BibitemShut  [1]{\csname bibitem#1\endcsname}%
\let\auto@bib@innerbib\@empty
\bibitem [{\citenamefont {Kimble}(2008)}]{Kimble2008}%
  \BibitemOpen
  \bibfield  {author} {\bibinfo {author} {\bibfnamefont {H.~J.}\ \bibnamefont
  {Kimble}},\ }\bibfield  {title} {\bibinfo {title} {{The quantum internet}},\
  }\href@noop {} {\bibfield  {journal} {\bibinfo  {journal} {Nature}\ }\textbf
  {\bibinfo {volume} {453}},\ \bibinfo {pages} {1023} (\bibinfo {year}
  {2008})}\BibitemShut {NoStop}%
\bibitem [{\citenamefont {Simon}(2017)}]{Simon2017}%
  \BibitemOpen
  \bibfield  {author} {\bibinfo {author} {\bibfnamefont {C.}~\bibnamefont
  {Simon}},\ }\bibfield  {title} {\bibinfo {title} {{Towards a global quantum
  network}},\ }\href@noop {} {\bibfield  {journal} {\bibinfo  {journal} {Nature
  Photonics}\ }\textbf {\bibinfo {volume} {11}},\ \bibinfo {pages} {678}
  (\bibinfo {year} {2017})}\BibitemShut {NoStop}%
\bibitem [{\citenamefont {Wehner}\ \emph {et~al.}(2018)\citenamefont {Wehner},
  \citenamefont {Elkouss},\ and\ \citenamefont {Hanson}}]{Wehner2018}%
  \BibitemOpen
  \bibfield  {author} {\bibinfo {author} {\bibfnamefont {S.}~\bibnamefont
  {Wehner}}, \bibinfo {author} {\bibfnamefont {D.}~\bibnamefont {Elkouss}},\
  and\ \bibinfo {author} {\bibfnamefont {R.}~\bibnamefont {Hanson}},\
  }\bibfield  {title} {\bibinfo {title} {{Quantum internet: A vision for the
  road ahead}},\ }\href@noop {} {\bibfield  {journal} {\bibinfo  {journal}
  {Science}\ }\textbf {\bibinfo {volume} {362}},\ \bibinfo {pages} {eaam9288}
  (\bibinfo {year} {2018})}\BibitemShut {NoStop}%
\bibitem [{\citenamefont {Sangouard}\ \emph {et~al.}(2011)\citenamefont
  {Sangouard}, \citenamefont {Simon}, \citenamefont {de~Riedmatten},\ and\
  \citenamefont {Gisin}}]{Sangouard2011}%
  \BibitemOpen
  \bibfield  {author} {\bibinfo {author} {\bibfnamefont {N.}~\bibnamefont
  {Sangouard}}, \bibinfo {author} {\bibfnamefont {C.}~\bibnamefont {Simon}},
  \bibinfo {author} {\bibfnamefont {H.}~\bibnamefont {de~Riedmatten}},\ and\
  \bibinfo {author} {\bibfnamefont {N.}~\bibnamefont {Gisin}},\ }\bibfield
  {title} {\bibinfo {title} {{Quantum repeaters based on atomic ensembles and
  linear optics}},\ }\href@noop {} {\bibfield  {journal} {\bibinfo  {journal}
  {Reviews of Modern Physics}\ }\textbf {\bibinfo {volume} {83}},\ \bibinfo
  {pages} {33} (\bibinfo {year} {2011})}\BibitemShut {NoStop}%
\bibitem [{\citenamefont {Duan}\ \emph {et~al.}(2001)\citenamefont {Duan},
  \citenamefont {Lukin}, \citenamefont {Cirac},\ and\ \citenamefont
  {Zoller}}]{Duan2001}%
  \BibitemOpen
  \bibfield  {author} {\bibinfo {author} {\bibfnamefont {L.-M.}\ \bibnamefont
  {Duan}}, \bibinfo {author} {\bibfnamefont {M.}~\bibnamefont {Lukin}},
  \bibinfo {author} {\bibfnamefont {J.~I.}\ \bibnamefont {Cirac}},\ and\
  \bibinfo {author} {\bibfnamefont {P.}~\bibnamefont {Zoller}},\ }\bibfield
  {title} {\bibinfo {title} {Long-distance quantum communication with atomic
  ensembles and linear optics},\ }\href@noop {} {\bibfield  {journal} {\bibinfo
   {journal} {Nature}\ }\textbf {\bibinfo {volume} {414}},\ \bibinfo {pages}
  {413} (\bibinfo {year} {2001})}\BibitemShut {NoStop}%
\bibitem [{\citenamefont {Hammerer}\ \emph {et~al.}(2010)\citenamefont
  {Hammerer}, \citenamefont {S{\o}rensen},\ and\ \citenamefont
  {Polzik}}]{Hammerer2010}%
  \BibitemOpen
  \bibfield  {author} {\bibinfo {author} {\bibfnamefont {K.}~\bibnamefont
  {Hammerer}}, \bibinfo {author} {\bibfnamefont {A.~S.}\ \bibnamefont
  {S{\o}rensen}},\ and\ \bibinfo {author} {\bibfnamefont {E.~S.}\ \bibnamefont
  {Polzik}},\ }\bibfield  {title} {\bibinfo {title} {{Quantum interface between
  light and atomic ensembles}},\ }\href@noop {} {\bibfield  {journal} {\bibinfo
   {journal} {Reviews of Modern Physics}\ }\textbf {\bibinfo {volume} {82}},\
  \bibinfo {pages} {1041} (\bibinfo {year} {2010})}\BibitemShut {NoStop}%
\bibitem [{\citenamefont {Simon}\ \emph {et~al.}(2007)\citenamefont {Simon},
  \citenamefont {Tanji}, \citenamefont {Thompson},\ and\ \citenamefont
  {Vuleti{\'{c}}}}]{Simon2007}%
  \BibitemOpen
  \bibfield  {author} {\bibinfo {author} {\bibfnamefont {J.}~\bibnamefont
  {Simon}}, \bibinfo {author} {\bibfnamefont {H.}~\bibnamefont {Tanji}},
  \bibinfo {author} {\bibfnamefont {J.~K.}\ \bibnamefont {Thompson}},\ and\
  \bibinfo {author} {\bibfnamefont {V.}~\bibnamefont {Vuleti{\'{c}}}},\
  }\bibfield  {title} {\bibinfo {title} {{Interfacing Collective Atomic
  Excitations and Single Photons}},\ }\href@noop {} {\bibfield  {journal}
  {\bibinfo  {journal} {Physical Review Letters}\ }\textbf {\bibinfo {volume}
  {98}},\ \bibinfo {pages} {183601} (\bibinfo {year} {2007})}\BibitemShut
  {NoStop}%
\bibitem [{\citenamefont {Zhao}\ \emph
  {et~al.}(2009{\natexlab{a}})\citenamefont {Zhao}, \citenamefont {Chen},
  \citenamefont {Bao}, \citenamefont {Strassel}, \citenamefont {Chuu},
  \citenamefont {Jin}, \citenamefont {Schmiedmayer}, \citenamefont {Yuan},
  \citenamefont {Chen},\ and\ \citenamefont {Pan}}]{Zhao2009}%
  \BibitemOpen
  \bibfield  {author} {\bibinfo {author} {\bibfnamefont {B.}~\bibnamefont
  {Zhao}}, \bibinfo {author} {\bibfnamefont {Y.-A.}\ \bibnamefont {Chen}},
  \bibinfo {author} {\bibfnamefont {X.-H.}\ \bibnamefont {Bao}}, \bibinfo
  {author} {\bibfnamefont {T.}~\bibnamefont {Strassel}}, \bibinfo {author}
  {\bibfnamefont {C.-S.}\ \bibnamefont {Chuu}}, \bibinfo {author}
  {\bibfnamefont {X.-M.}\ \bibnamefont {Jin}}, \bibinfo {author} {\bibfnamefont
  {J.}~\bibnamefont {Schmiedmayer}}, \bibinfo {author} {\bibfnamefont {Z.-S.}\
  \bibnamefont {Yuan}}, \bibinfo {author} {\bibfnamefont {S.}~\bibnamefont
  {Chen}},\ and\ \bibinfo {author} {\bibfnamefont {J.-W.}\ \bibnamefont
  {Pan}},\ }\bibfield  {title} {\bibinfo {title} {{A millisecond quantum memory
  for scalable quantum networks}},\ }\href@noop {} {\bibfield  {journal}
  {\bibinfo  {journal} {Nature Physics}\ }\textbf {\bibinfo {volume} {5}},\
  \bibinfo {pages} {95} (\bibinfo {year} {2009}{\natexlab{a}})}\BibitemShut
  {NoStop}%
\bibitem [{\citenamefont {Zhao}\ \emph
  {et~al.}(2009{\natexlab{b}})\citenamefont {Zhao}, \citenamefont {Dudin},
  \citenamefont {Jenkins}, \citenamefont {Campbell}, \citenamefont
  {Matsukevich}, \citenamefont {Kennedy},\ and\ \citenamefont
  {Kuzmich}}]{zhao2009long}%
  \BibitemOpen
  \bibfield  {author} {\bibinfo {author} {\bibfnamefont {R.}~\bibnamefont
  {Zhao}}, \bibinfo {author} {\bibfnamefont {Y.}~\bibnamefont {Dudin}},
  \bibinfo {author} {\bibfnamefont {S.}~\bibnamefont {Jenkins}}, \bibinfo
  {author} {\bibfnamefont {C.}~\bibnamefont {Campbell}}, \bibinfo {author}
  {\bibfnamefont {D.}~\bibnamefont {Matsukevich}}, \bibinfo {author}
  {\bibfnamefont {T.}~\bibnamefont {Kennedy}},\ and\ \bibinfo {author}
  {\bibfnamefont {A.}~\bibnamefont {Kuzmich}},\ }\bibfield  {title} {\bibinfo
  {title} {Long-lived quantum memory},\ }\href@noop {} {\bibfield  {journal}
  {\bibinfo  {journal} {Nature Physics}\ }\textbf {\bibinfo {volume} {5}},\
  \bibinfo {pages} {100} (\bibinfo {year} {2009}{\natexlab{b}})}\BibitemShut
  {NoStop}%
\bibitem [{\citenamefont {Lan}\ \emph {et~al.}(2009)\citenamefont {Lan},
  \citenamefont {Radnaev}, \citenamefont {Collins}, \citenamefont
  {Matsukevich}, \citenamefont {Kennedy},\ and\ \citenamefont
  {Kuzmich}}]{Lan2009}%
  \BibitemOpen
  \bibfield  {author} {\bibinfo {author} {\bibfnamefont {S.-Y.}\ \bibnamefont
  {Lan}}, \bibinfo {author} {\bibfnamefont {A.~G.}\ \bibnamefont {Radnaev}},
  \bibinfo {author} {\bibfnamefont {O.~A.}\ \bibnamefont {Collins}}, \bibinfo
  {author} {\bibfnamefont {D.~N.}\ \bibnamefont {Matsukevich}}, \bibinfo
  {author} {\bibfnamefont {T.~A.}\ \bibnamefont {Kennedy}},\ and\ \bibinfo
  {author} {\bibfnamefont {A.}~\bibnamefont {Kuzmich}},\ }\bibfield  {title}
  {\bibinfo {title} {A multiplexed quantum memory},\ }\href@noop {} {\bibfield
  {journal} {\bibinfo  {journal} {Optics Express}\ }\textbf {\bibinfo {volume}
  {17}},\ \bibinfo {pages} {13639} (\bibinfo {year} {2009})}\BibitemShut
  {NoStop}%
\bibitem [{\citenamefont {Dudin}\ \emph {et~al.}(2010)\citenamefont {Dudin},
  \citenamefont {Radnaev}, \citenamefont {Zhao}, \citenamefont {Blumoff},
  \citenamefont {Kennedy},\ and\ \citenamefont {Kuzmich}}]{Dudin2010}%
  \BibitemOpen
  \bibfield  {author} {\bibinfo {author} {\bibfnamefont {Y.~O.}\ \bibnamefont
  {Dudin}}, \bibinfo {author} {\bibfnamefont {A.~G.}\ \bibnamefont {Radnaev}},
  \bibinfo {author} {\bibfnamefont {R.}~\bibnamefont {Zhao}}, \bibinfo {author}
  {\bibfnamefont {J.~Z.}\ \bibnamefont {Blumoff}}, \bibinfo {author}
  {\bibfnamefont {T.~A.~B.}\ \bibnamefont {Kennedy}},\ and\ \bibinfo {author}
  {\bibfnamefont {A.}~\bibnamefont {Kuzmich}},\ }\bibfield  {title} {\bibinfo
  {title} {{Entanglement of Light-Shift Compensated Atomic Spin Waves with
  Telecom Light}},\ }\href@noop {} {\bibfield  {journal} {\bibinfo  {journal}
  {Physical Review Letters}\ }\textbf {\bibinfo {volume} {105}},\ \bibinfo
  {pages} {260502} (\bibinfo {year} {2010})}\BibitemShut {NoStop}%
\bibitem [{\citenamefont {Radnaev}\ \emph {et~al.}(2010)\citenamefont
  {Radnaev}, \citenamefont {Dudin}, \citenamefont {Zhao}, \citenamefont {Jen},
  \citenamefont {Jenkins}, \citenamefont {Kuzmich},\ and\ \citenamefont
  {Kennedy}}]{Radnaev2010}%
  \BibitemOpen
  \bibfield  {author} {\bibinfo {author} {\bibfnamefont {A.~G.}\ \bibnamefont
  {Radnaev}}, \bibinfo {author} {\bibfnamefont {Y.~O.}\ \bibnamefont {Dudin}},
  \bibinfo {author} {\bibfnamefont {R.}~\bibnamefont {Zhao}}, \bibinfo {author}
  {\bibfnamefont {H.~H.}\ \bibnamefont {Jen}}, \bibinfo {author} {\bibfnamefont
  {S.~D.}\ \bibnamefont {Jenkins}}, \bibinfo {author} {\bibfnamefont
  {A.}~\bibnamefont {Kuzmich}},\ and\ \bibinfo {author} {\bibfnamefont
  {T.~A.~B.}\ \bibnamefont {Kennedy}},\ }\bibfield  {title} {\bibinfo {title}
  {A quantum memory with telecom-wavelength conversion},\ }\href@noop {}
  {\bibfield  {journal} {\bibinfo  {journal} {Nature Physics}\ }\textbf
  {\bibinfo {volume} {6}},\ \bibinfo {pages} {894} (\bibinfo {year}
  {2010})}\BibitemShut {NoStop}%
\bibitem [{\citenamefont {Bao}\ \emph {et~al.}(2012)\citenamefont {Bao},
  \citenamefont {Reingruber}, \citenamefont {Dietrich}, \citenamefont {Rui},
  \citenamefont {D\"{u}ck}, \citenamefont {Strassel}, \citenamefont {Li},
  \citenamefont {Liu}, \citenamefont {Zhao},\ and\ \citenamefont
  {Pan}}]{Bao2012}%
  \BibitemOpen
  \bibfield  {author} {\bibinfo {author} {\bibfnamefont {X.-H.}\ \bibnamefont
  {Bao}}, \bibinfo {author} {\bibfnamefont {A.}~\bibnamefont {Reingruber}},
  \bibinfo {author} {\bibfnamefont {P.}~\bibnamefont {Dietrich}}, \bibinfo
  {author} {\bibfnamefont {J.}~\bibnamefont {Rui}}, \bibinfo {author}
  {\bibfnamefont {A.}~\bibnamefont {D\"{u}ck}}, \bibinfo {author}
  {\bibfnamefont {T.}~\bibnamefont {Strassel}}, \bibinfo {author}
  {\bibfnamefont {L.}~\bibnamefont {Li}}, \bibinfo {author} {\bibfnamefont
  {N.-L.}\ \bibnamefont {Liu}}, \bibinfo {author} {\bibfnamefont
  {B.}~\bibnamefont {Zhao}},\ and\ \bibinfo {author} {\bibfnamefont {J.-W.}\
  \bibnamefont {Pan}},\ }\bibfield  {title} {\bibinfo {title} {{Efficient and
  long-lived quantum memory with cold atoms inside a ring cavity}},\
  }\href@noop {} {\bibfield  {journal} {\bibinfo  {journal} {Nature Physics}\
  }\textbf {\bibinfo {volume} {8}},\ \bibinfo {pages} {517} (\bibinfo {year}
  {2012})}\BibitemShut {NoStop}%
\bibitem [{\citenamefont {Dudin}\ \emph {et~al.}(2013)\citenamefont {Dudin},
  \citenamefont {Li},\ and\ \citenamefont {Kuzmich}}]{Dudin2013}%
  \BibitemOpen
  \bibfield  {author} {\bibinfo {author} {\bibfnamefont {Y.}~\bibnamefont
  {Dudin}}, \bibinfo {author} {\bibfnamefont {L.}~\bibnamefont {Li}},\ and\
  \bibinfo {author} {\bibfnamefont {A.}~\bibnamefont {Kuzmich}},\ }\bibfield
  {title} {\bibinfo {title} {{Light storage on the time scale of a minute}},\
  }\href@noop {} {\bibfield  {journal} {\bibinfo  {journal} {Physical Review
  A}\ }\textbf {\bibinfo {volume} {87}},\ \bibinfo {pages} {31801} (\bibinfo
  {year} {2013})}\BibitemShut {NoStop}%
\bibitem [{\citenamefont {Xu}\ \emph {et~al.}(2013)\citenamefont {Xu},
  \citenamefont {Wu}, \citenamefont {Tian}, \citenamefont {Chen}, \citenamefont
  {Zhang}, \citenamefont {Yan}, \citenamefont {Li}, \citenamefont {Wang},
  \citenamefont {Xie},\ and\ \citenamefont {Peng}}]{xu_long_2013}%
  \BibitemOpen
  \bibfield  {author} {\bibinfo {author} {\bibfnamefont {Z.}~\bibnamefont
  {Xu}}, \bibinfo {author} {\bibfnamefont {Y.}~\bibnamefont {Wu}}, \bibinfo
  {author} {\bibfnamefont {L.}~\bibnamefont {Tian}}, \bibinfo {author}
  {\bibfnamefont {L.}~\bibnamefont {Chen}}, \bibinfo {author} {\bibfnamefont
  {Z.}~\bibnamefont {Zhang}}, \bibinfo {author} {\bibfnamefont
  {Z.}~\bibnamefont {Yan}}, \bibinfo {author} {\bibfnamefont {S.}~\bibnamefont
  {Li}}, \bibinfo {author} {\bibfnamefont {H.}~\bibnamefont {Wang}}, \bibinfo
  {author} {\bibfnamefont {C.}~\bibnamefont {Xie}},\ and\ \bibinfo {author}
  {\bibfnamefont {K.}~\bibnamefont {Peng}},\ }\bibfield  {title} {\bibinfo
  {title} {Long {Lifetime} and {High}-{Fidelity} {Quantum} {Memory} of
  {Photonic} {Polarization} {Qubit} by {Lifting} {Zeeman} {Degeneracy}},\
  }\href@noop {} {\bibfield  {journal} {\bibinfo  {journal} {Physical Review
  Letters}\ }\textbf {\bibinfo {volume} {111}},\ \bibinfo {pages} {240503}
  (\bibinfo {year} {2013})}\BibitemShut {NoStop}%
\bibitem [{\citenamefont {Cho}\ \emph {et~al.}(2016)\citenamefont {Cho},
  \citenamefont {Campbell}, \citenamefont {Everett}, \citenamefont {Bernu},
  \citenamefont {Higginbottom}, \citenamefont {Cao}, \citenamefont {Geng},
  \citenamefont {Robins}, \citenamefont {Lam},\ and\ \citenamefont
  {Buchler}}]{cho_highly_2016}%
  \BibitemOpen
  \bibfield  {author} {\bibinfo {author} {\bibfnamefont {Y.-W.}\ \bibnamefont
  {Cho}}, \bibinfo {author} {\bibfnamefont {G.~T.}\ \bibnamefont {Campbell}},
  \bibinfo {author} {\bibfnamefont {J.~L.}\ \bibnamefont {Everett}}, \bibinfo
  {author} {\bibfnamefont {J.}~\bibnamefont {Bernu}}, \bibinfo {author}
  {\bibfnamefont {D.~B.}\ \bibnamefont {Higginbottom}}, \bibinfo {author}
  {\bibfnamefont {M.~T.}\ \bibnamefont {Cao}}, \bibinfo {author} {\bibfnamefont
  {J.}~\bibnamefont {Geng}}, \bibinfo {author} {\bibfnamefont {N.~P.}\
  \bibnamefont {Robins}}, \bibinfo {author} {\bibfnamefont {P.~K.}\
  \bibnamefont {Lam}},\ and\ \bibinfo {author} {\bibfnamefont {B.~C.}\
  \bibnamefont {Buchler}},\ }\bibfield  {title} {\bibinfo {title} {Highly
  efficient optical quantum memory with long coherence time in cold atoms},\
  }\href@noop {} {\bibfield  {journal} {\bibinfo  {journal} {Optica}\ }\textbf
  {\bibinfo {volume} {3}},\ \bibinfo {pages} {100} (\bibinfo {year}
  {2016})}\BibitemShut {NoStop}%
\bibitem [{\citenamefont {Yang}\ \emph {et~al.}(2016)\citenamefont {Yang},
  \citenamefont {Wang}, \citenamefont {Bao},\ and\ \citenamefont
  {Pan}}]{Yang2016}%
  \BibitemOpen
  \bibfield  {author} {\bibinfo {author} {\bibfnamefont {S.-J.}\ \bibnamefont
  {Yang}}, \bibinfo {author} {\bibfnamefont {X.-J.}\ \bibnamefont {Wang}},
  \bibinfo {author} {\bibfnamefont {X.-H.}\ \bibnamefont {Bao}},\ and\ \bibinfo
  {author} {\bibfnamefont {J.-W.}\ \bibnamefont {Pan}},\ }\bibfield  {title}
  {\bibinfo {title} {An efficient quantum light--matter interface with
  sub-second lifetime},\ }\href@noop {} {\bibfield  {journal} {\bibinfo
  {journal} {Nature Photonics}\ }\textbf {\bibinfo {volume} {10}},\ \bibinfo
  {pages} {381} (\bibinfo {year} {2016})}\BibitemShut {NoStop}%
\bibitem [{\citenamefont {Pu}\ \emph {et~al.}(2017)\citenamefont {Pu},
  \citenamefont {Jiang}, \citenamefont {Chang}, \citenamefont {Yang},
  \citenamefont {Li},\ and\ \citenamefont {Duan}}]{Pu2017}%
  \BibitemOpen
  \bibfield  {author} {\bibinfo {author} {\bibfnamefont {Y.-F.}\ \bibnamefont
  {Pu}}, \bibinfo {author} {\bibfnamefont {N.}~\bibnamefont {Jiang}}, \bibinfo
  {author} {\bibfnamefont {W.}~\bibnamefont {Chang}}, \bibinfo {author}
  {\bibfnamefont {H.-X.}\ \bibnamefont {Yang}}, \bibinfo {author}
  {\bibfnamefont {C.}~\bibnamefont {Li}},\ and\ \bibinfo {author}
  {\bibfnamefont {L.-M.}\ \bibnamefont {Duan}},\ }\bibfield  {title} {\bibinfo
  {title} {{Experimental realization of a multiplexed quantum memory with 225
  individually accessible memory cells}},\ }\href@noop {} {\bibfield  {journal}
  {\bibinfo  {journal} {Nature Communications}\ }\textbf {\bibinfo {volume}
  {8}},\ \bibinfo {pages} {15359} (\bibinfo {year} {2017})}\BibitemShut
  {NoStop}%
\bibitem [{\citenamefont {Tian}\ \emph {et~al.}(2017)\citenamefont {Tian},
  \citenamefont {Xu}, \citenamefont {Chen}, \citenamefont {Ge}, \citenamefont
  {Yuan}, \citenamefont {Wen}, \citenamefont {Wang}, \citenamefont {Li},\ and\
  \citenamefont {Wang}}]{tian_spatial_2017}%
  \BibitemOpen
  \bibfield  {author} {\bibinfo {author} {\bibfnamefont {L.}~\bibnamefont
  {Tian}}, \bibinfo {author} {\bibfnamefont {Z.}~\bibnamefont {Xu}}, \bibinfo
  {author} {\bibfnamefont {L.}~\bibnamefont {Chen}}, \bibinfo {author}
  {\bibfnamefont {W.}~\bibnamefont {Ge}}, \bibinfo {author} {\bibfnamefont
  {H.}~\bibnamefont {Yuan}}, \bibinfo {author} {\bibfnamefont {Y.}~\bibnamefont
  {Wen}}, \bibinfo {author} {\bibfnamefont {S.}~\bibnamefont {Wang}}, \bibinfo
  {author} {\bibfnamefont {S.}~\bibnamefont {Li}},\ and\ \bibinfo {author}
  {\bibfnamefont {H.}~\bibnamefont {Wang}},\ }\bibfield  {title} {\bibinfo
  {title} {Spatial {Multiplexing} of {Atom}-{Photon} {Entanglement} {Sources}
  using {Feedforward} {Control} and {Switching} {Networks}},\ }\href@noop {}
  {\bibfield  {journal} {\bibinfo  {journal} {Physical Review Letters}\
  }\textbf {\bibinfo {volume} {119}},\ \bibinfo {pages} {130505} (\bibinfo
  {year} {2017})}\BibitemShut {NoStop}%
\bibitem [{\citenamefont {Chrapkiewicz}\ \emph {et~al.}(2017)\citenamefont
  {Chrapkiewicz}, \citenamefont {Dabrowski},\ and\ \citenamefont
  {Wasilewski}}]{chrapkiewicz_high-capacity_2017}%
  \BibitemOpen
  \bibfield  {author} {\bibinfo {author} {\bibfnamefont {R.}~\bibnamefont
  {Chrapkiewicz}}, \bibinfo {author} {\bibfnamefont {M.}~\bibnamefont
  {Dabrowski}},\ and\ \bibinfo {author} {\bibfnamefont {W.}~\bibnamefont
  {Wasilewski}},\ }\bibfield  {title} {\bibinfo {title} {High-capacity
  angularly multiplexed holographic memory operating at the single-photon
  level},\ }\href@noop {} {\bibfield  {journal} {\bibinfo  {journal} {Physical
  Review Letters}\ }\textbf {\bibinfo {volume} {118}},\ \bibinfo {pages}
  {063603} (\bibinfo {year} {2017})}\BibitemShut {NoStop}%
\bibitem [{\citenamefont {Shi}\ \emph {et~al.}(2018)\citenamefont {Shi},
  \citenamefont {Ding},\ and\ \citenamefont {Zhang}}]{Shi2018}%
  \BibitemOpen
  \bibfield  {author} {\bibinfo {author} {\bibfnamefont {B.-S.}\ \bibnamefont
  {Shi}}, \bibinfo {author} {\bibfnamefont {D.-S.}\ \bibnamefont {Ding}},\ and\
  \bibinfo {author} {\bibfnamefont {W.}~\bibnamefont {Zhang}},\ }\bibfield
  {title} {\bibinfo {title} {Quantum storage of orbital angular momentum
  entanglement in cold atomic ensembles},\ }\href@noop {} {\bibfield  {journal}
  {\bibinfo  {journal} {Journal of Physics B: Atomic, Molecular and Optical
  Physics}\ }\textbf {\bibinfo {volume} {51}},\ \bibinfo {pages} {032004}
  (\bibinfo {year} {2018})}\BibitemShut {NoStop}%
\bibitem [{\citenamefont {Hsiao}\ \emph {et~al.}(2018)\citenamefont {Hsiao},
  \citenamefont {Tsai}, \citenamefont {Chen}, \citenamefont {Lin},
  \citenamefont {Hung}, \citenamefont {Lee}, \citenamefont {Chen},
  \citenamefont {Chen}, \citenamefont {Yu},\ and\ \citenamefont
  {Chen}}]{Hsiao2018}%
  \BibitemOpen
  \bibfield  {author} {\bibinfo {author} {\bibfnamefont {Y.-F.}\ \bibnamefont
  {Hsiao}}, \bibinfo {author} {\bibfnamefont {P.-J.}\ \bibnamefont {Tsai}},
  \bibinfo {author} {\bibfnamefont {H.-S.}\ \bibnamefont {Chen}}, \bibinfo
  {author} {\bibfnamefont {S.-X.}\ \bibnamefont {Lin}}, \bibinfo {author}
  {\bibfnamefont {C.-C.}\ \bibnamefont {Hung}}, \bibinfo {author}
  {\bibfnamefont {C.-H.}\ \bibnamefont {Lee}}, \bibinfo {author} {\bibfnamefont
  {Y.-H.}\ \bibnamefont {Chen}}, \bibinfo {author} {\bibfnamefont {Y.-F.}\
  \bibnamefont {Chen}}, \bibinfo {author} {\bibfnamefont {I.~A.}\ \bibnamefont
  {Yu}},\ and\ \bibinfo {author} {\bibfnamefont {Y.-C.}\ \bibnamefont {Chen}},\
  }\bibfield  {title} {\bibinfo {title} {{Highly Efficient Coherent Optical
  Memory Based on Electromagnetically Induced Transparency}},\ }\href@noop {}
  {\bibfield  {journal} {\bibinfo  {journal} {Physical Review Letters}\
  }\textbf {\bibinfo {volume} {120}},\ \bibinfo {pages} {183602} (\bibinfo
  {year} {2018})}\BibitemShut {NoStop}%
\bibitem [{\citenamefont {Vernaz-Gris}\ \emph {et~al.}(2018)\citenamefont
  {Vernaz-Gris}, \citenamefont {Huang}, \citenamefont {Cao}, \citenamefont
  {Sheremet},\ and\ \citenamefont {Laurat}}]{vernaz-gris2018}%
  \BibitemOpen
  \bibfield  {author} {\bibinfo {author} {\bibfnamefont {P.}~\bibnamefont
  {Vernaz-Gris}}, \bibinfo {author} {\bibfnamefont {K.}~\bibnamefont {Huang}},
  \bibinfo {author} {\bibfnamefont {M.}~\bibnamefont {Cao}}, \bibinfo {author}
  {\bibfnamefont {A.~S.}\ \bibnamefont {Sheremet}},\ and\ \bibinfo {author}
  {\bibfnamefont {J.}~\bibnamefont {Laurat}},\ }\bibfield  {title} {\bibinfo
  {title} {Highly-efficient quantum memory for polarization qubits in a
  spatially-multiplexed cold atomic ensemble},\ }\href@noop {} {\bibfield
  {journal} {\bibinfo  {journal} {Nature Communications}\ }\textbf {\bibinfo
  {volume} {9}},\ \bibinfo {pages} {363} (\bibinfo {year} {2018})}\BibitemShut
  {NoStop}%
\bibitem [{\citenamefont {Wang}\ \emph {et~al.}(2019)\citenamefont {Wang},
  \citenamefont {Li}, \citenamefont {Zhang}, \citenamefont {Su}, \citenamefont
  {Zhou}, \citenamefont {Liao}, \citenamefont {Du}, \citenamefont {Yan},\ and\
  \citenamefont {Zhu}}]{Wang2019}%
  \BibitemOpen
  \bibfield  {author} {\bibinfo {author} {\bibfnamefont {Y.}~\bibnamefont
  {Wang}}, \bibinfo {author} {\bibfnamefont {J.}~\bibnamefont {Li}}, \bibinfo
  {author} {\bibfnamefont {S.}~\bibnamefont {Zhang}}, \bibinfo {author}
  {\bibfnamefont {K.}~\bibnamefont {Su}}, \bibinfo {author} {\bibfnamefont
  {Y.}~\bibnamefont {Zhou}}, \bibinfo {author} {\bibfnamefont {K.}~\bibnamefont
  {Liao}}, \bibinfo {author} {\bibfnamefont {S.}~\bibnamefont {Du}}, \bibinfo
  {author} {\bibfnamefont {H.}~\bibnamefont {Yan}},\ and\ \bibinfo {author}
  {\bibfnamefont {S.-L.}\ \bibnamefont {Zhu}},\ }\bibfield  {title} {\bibinfo
  {title} {{Efficient quantum memory for single-photon polarization qubits}},\
  }\href@noop {} {\bibfield  {journal} {\bibinfo  {journal} {Nature Photonics}\
  ,\ \bibinfo {pages} {22}} (\bibinfo {year} {2019})}\BibitemShut {NoStop}%
\bibitem [{\citenamefont {Chang}\ \emph {et~al.}(2019)\citenamefont {Chang},
  \citenamefont {Li}, \citenamefont {Wu}, \citenamefont {Jiang}, \citenamefont
  {Zhang}, \citenamefont {Pu}, \citenamefont {Chang},\ and\ \citenamefont
  {Duan}}]{chang_long-distance_2019}%
  \BibitemOpen
  \bibfield  {author} {\bibinfo {author} {\bibfnamefont {W.}~\bibnamefont
  {Chang}}, \bibinfo {author} {\bibfnamefont {C.}~\bibnamefont {Li}}, \bibinfo
  {author} {\bibfnamefont {Y.-K.}\ \bibnamefont {Wu}}, \bibinfo {author}
  {\bibfnamefont {N.}~\bibnamefont {Jiang}}, \bibinfo {author} {\bibfnamefont
  {S.}~\bibnamefont {Zhang}}, \bibinfo {author} {\bibfnamefont {Y.-F.}\
  \bibnamefont {Pu}}, \bibinfo {author} {\bibfnamefont {X.-Y.}\ \bibnamefont
  {Chang}},\ and\ \bibinfo {author} {\bibfnamefont {L.-M.}\ \bibnamefont
  {Duan}},\ }\bibfield  {title} {\bibinfo {title} {Long-{Distance}
  {Entanglement} between a {Multiplexed} {Quantum} {Memory} and a {Telecom}
  {Photon}},\ }\href@noop {} {\bibfield  {journal} {\bibinfo  {journal}
  {Physical Review X}\ }\textbf {\bibinfo {volume} {9}},\ \bibinfo {pages}
  {041033} (\bibinfo {year} {2019})}\BibitemShut {NoStop}%
\bibitem [{\citenamefont {Jing}\ \emph {et~al.}(2019)\citenamefont {Jing},
  \citenamefont {Wang}, \citenamefont {Yu}, \citenamefont {Sun}, \citenamefont
  {Jiang}, \citenamefont {Yang}, \citenamefont {Jiang}, \citenamefont {Luo},
  \citenamefont {Zhang}, \citenamefont {Jiang}, \citenamefont {Bao},\ and\
  \citenamefont {Pan}}]{jing_entanglement_2019}%
  \BibitemOpen
  \bibfield  {author} {\bibinfo {author} {\bibfnamefont {B.}~\bibnamefont
  {Jing}}, \bibinfo {author} {\bibfnamefont {X.-J.}\ \bibnamefont {Wang}},
  \bibinfo {author} {\bibfnamefont {Y.}~\bibnamefont {Yu}}, \bibinfo {author}
  {\bibfnamefont {P.-F.}\ \bibnamefont {Sun}}, \bibinfo {author} {\bibfnamefont
  {Y.}~\bibnamefont {Jiang}}, \bibinfo {author} {\bibfnamefont {S.-J.}\
  \bibnamefont {Yang}}, \bibinfo {author} {\bibfnamefont {W.-H.}\ \bibnamefont
  {Jiang}}, \bibinfo {author} {\bibfnamefont {X.-Y.}\ \bibnamefont {Luo}},
  \bibinfo {author} {\bibfnamefont {J.}~\bibnamefont {Zhang}}, \bibinfo
  {author} {\bibfnamefont {X.}~\bibnamefont {Jiang}}, \bibinfo {author}
  {\bibfnamefont {X.-H.}\ \bibnamefont {Bao}},\ and\ \bibinfo {author}
  {\bibfnamefont {J.-W.}\ \bibnamefont {Pan}},\ }\bibfield  {title} {\bibinfo
  {title} {Entanglement of three quantum memories via interference of three
  single photons},\ }\href@noop {} {\bibfield  {journal} {\bibinfo  {journal}
  {Nature Photonics}\ }\textbf {\bibinfo {volume} {13}},\ \bibinfo {pages}
  {210} (\bibinfo {year} {2019})}\BibitemShut {NoStop}%
\bibitem [{\citenamefont {Cao}\ \emph {et~al.}(2020)\citenamefont {Cao},
  \citenamefont {Hoffet}, \citenamefont {Qiu}, \citenamefont {Sheremet},\ and\
  \citenamefont {Laurat}}]{cao_efficient_2020}%
  \BibitemOpen
  \bibfield  {author} {\bibinfo {author} {\bibfnamefont {M.}~\bibnamefont
  {Cao}}, \bibinfo {author} {\bibfnamefont {F.}~\bibnamefont {Hoffet}},
  \bibinfo {author} {\bibfnamefont {S.}~\bibnamefont {Qiu}}, \bibinfo {author}
  {\bibfnamefont {A.~S.}\ \bibnamefont {Sheremet}},\ and\ \bibinfo {author}
  {\bibfnamefont {J.}~\bibnamefont {Laurat}},\ }\bibfield  {title} {\bibinfo
  {title} {Efficient reversible entanglement transfer between light and quantum
  memories},\ }\href@noop {} {\bibfield  {journal} {\bibinfo  {journal}
  {Optica}\ }\textbf {\bibinfo {volume} {7}},\ \bibinfo {pages} {1440}
  (\bibinfo {year} {2020})}\BibitemShut {NoStop}%
\bibitem [{\citenamefont {Yu}\ \emph {et~al.}(2020)\citenamefont {Yu},
  \citenamefont {Ma}, \citenamefont {Luo}, \citenamefont {Jing}, \citenamefont
  {Sun}, \citenamefont {Fang}, \citenamefont {Yang}, \citenamefont {Liu},
  \citenamefont {Zheng}, \citenamefont {Xie}, \citenamefont {Zhang},
  \citenamefont {You}, \citenamefont {Wang}, \citenamefont {Chen},
  \citenamefont {Zhang}, \citenamefont {Bao},\ and\ \citenamefont
  {Pan}}]{yu_entanglement_2020}%
  \BibitemOpen
  \bibfield  {author} {\bibinfo {author} {\bibfnamefont {Y.}~\bibnamefont
  {Yu}}, \bibinfo {author} {\bibfnamefont {F.}~\bibnamefont {Ma}}, \bibinfo
  {author} {\bibfnamefont {X.-Y.}\ \bibnamefont {Luo}}, \bibinfo {author}
  {\bibfnamefont {B.}~\bibnamefont {Jing}}, \bibinfo {author} {\bibfnamefont
  {P.-F.}\ \bibnamefont {Sun}}, \bibinfo {author} {\bibfnamefont {R.-Z.}\
  \bibnamefont {Fang}}, \bibinfo {author} {\bibfnamefont {C.-W.}\ \bibnamefont
  {Yang}}, \bibinfo {author} {\bibfnamefont {H.}~\bibnamefont {Liu}}, \bibinfo
  {author} {\bibfnamefont {M.-Y.}\ \bibnamefont {Zheng}}, \bibinfo {author}
  {\bibfnamefont {X.-P.}\ \bibnamefont {Xie}}, \bibinfo {author} {\bibfnamefont
  {W.-J.}\ \bibnamefont {Zhang}}, \bibinfo {author} {\bibfnamefont {L.-X.}\
  \bibnamefont {You}}, \bibinfo {author} {\bibfnamefont {Z.}~\bibnamefont
  {Wang}}, \bibinfo {author} {\bibfnamefont {T.-Y.}\ \bibnamefont {Chen}},
  \bibinfo {author} {\bibfnamefont {Q.}~\bibnamefont {Zhang}}, \bibinfo
  {author} {\bibfnamefont {X.-H.}\ \bibnamefont {Bao}},\ and\ \bibinfo {author}
  {\bibfnamefont {J.-W.}\ \bibnamefont {Pan}},\ }\bibfield  {title} {\bibinfo
  {title} {Entanglement of two quantum memories via fibres over dozens of
  kilometres},\ }\href@noop {} {\bibfield  {journal} {\bibinfo  {journal}
  {Nature}\ }\textbf {\bibinfo {volume} {578}},\ \bibinfo {pages} {240}
  (\bibinfo {year} {2020})}\BibitemShut {NoStop}%
\bibitem [{\citenamefont {Wang}\ \emph {et~al.}(2021)\citenamefont {Wang},
  \citenamefont {Yang}, \citenamefont {Sun}, \citenamefont {Jing},
  \citenamefont {Li}, \citenamefont {Zhou}, \citenamefont {Bao},\ and\
  \citenamefont {Pan}}]{wang_cavity-enhanced_2021}%
  \BibitemOpen
  \bibfield  {author} {\bibinfo {author} {\bibfnamefont {X.-J.}\ \bibnamefont
  {Wang}}, \bibinfo {author} {\bibfnamefont {S.-J.}\ \bibnamefont {Yang}},
  \bibinfo {author} {\bibfnamefont {P.-F.}\ \bibnamefont {Sun}}, \bibinfo
  {author} {\bibfnamefont {B.}~\bibnamefont {Jing}}, \bibinfo {author}
  {\bibfnamefont {J.}~\bibnamefont {Li}}, \bibinfo {author} {\bibfnamefont
  {M.-T.}\ \bibnamefont {Zhou}}, \bibinfo {author} {\bibfnamefont {X.-H.}\
  \bibnamefont {Bao}},\ and\ \bibinfo {author} {\bibfnamefont {J.-W.}\
  \bibnamefont {Pan}},\ }\bibfield  {title} {\bibinfo {title}
  {Cavity-{Enhanced} {Atom}-{Photon} {Entanglement} with {Subsecond}
  {Lifetime}},\ }\href@noop {} {\bibfield  {journal} {\bibinfo  {journal}
  {Physical Review Letters}\ }\textbf {\bibinfo {volume} {126}},\ \bibinfo
  {pages} {090501} (\bibinfo {year} {2021})}\BibitemShut {NoStop}%
\bibitem [{\citenamefont {Zhao}\ \emph {et~al.}(2010)\citenamefont {Zhao},
  \citenamefont {Müller}, \citenamefont {Hammerer}, \citenamefont {Zoller},\
  and\ \citenamefont {Muller}}]{Zhao2010c}%
  \BibitemOpen
  \bibfield  {author} {\bibinfo {author} {\bibfnamefont {B.}~\bibnamefont
  {Zhao}}, \bibinfo {author} {\bibfnamefont {M.}~\bibnamefont {Müller}},
  \bibinfo {author} {\bibfnamefont {K.}~\bibnamefont {Hammerer}}, \bibinfo
  {author} {\bibfnamefont {P.}~\bibnamefont {Zoller}},\ and\ \bibinfo {author}
  {\bibfnamefont {M.}~\bibnamefont {Muller}},\ }\bibfield  {title} {\bibinfo
  {title} {Efficient quantum repeater based on deterministic {Rydberg} gates},\
  }\href@noop {} {\bibfield  {journal} {\bibinfo  {journal} {Physical Review
  A}\ }\textbf {\bibinfo {volume} {81}},\ \bibinfo {pages} {052329} (\bibinfo
  {year} {2010})}\BibitemShut {NoStop}%
\bibitem [{\citenamefont {Han}\ \emph {et~al.}(2010)\citenamefont {Han},
  \citenamefont {He}, \citenamefont {Heshami}, \citenamefont {Li},\ and\
  \citenamefont {Simon}}]{Han2010}%
  \BibitemOpen
  \bibfield  {author} {\bibinfo {author} {\bibfnamefont {Y.}~\bibnamefont
  {Han}}, \bibinfo {author} {\bibfnamefont {B.}~\bibnamefont {He}}, \bibinfo
  {author} {\bibfnamefont {K.}~\bibnamefont {Heshami}}, \bibinfo {author}
  {\bibfnamefont {C.-Z.}\ \bibnamefont {Li}},\ and\ \bibinfo {author}
  {\bibfnamefont {C.}~\bibnamefont {Simon}},\ }\bibfield  {title} {\bibinfo
  {title} {Quantum repeaters based on {Rydberg}-blockade-coupled atomic
  ensembles},\ }\href@noop {} {\bibfield  {journal} {\bibinfo  {journal}
  {Physical Review A}\ }\textbf {\bibinfo {volume} {81}},\ \bibinfo {pages}
  {052311} (\bibinfo {year} {2010})}\BibitemShut {NoStop}%
\bibitem [{\citenamefont {Lukin}\ \emph {et~al.}(2001)\citenamefont {Lukin},
  \citenamefont {Fleischhauer}, \citenamefont {Cote}, \citenamefont {Duan},
  \citenamefont {Jaksch}, \citenamefont {Cirac},\ and\ \citenamefont
  {Zoller}}]{Lukin2001}%
  \BibitemOpen
  \bibfield  {author} {\bibinfo {author} {\bibfnamefont {M.~D.}\ \bibnamefont
  {Lukin}}, \bibinfo {author} {\bibfnamefont {M.}~\bibnamefont {Fleischhauer}},
  \bibinfo {author} {\bibfnamefont {R.}~\bibnamefont {Cote}}, \bibinfo {author}
  {\bibfnamefont {L.~M.}\ \bibnamefont {Duan}}, \bibinfo {author}
  {\bibfnamefont {D.}~\bibnamefont {Jaksch}}, \bibinfo {author} {\bibfnamefont
  {J.~I.}\ \bibnamefont {Cirac}},\ and\ \bibinfo {author} {\bibfnamefont
  {P.}~\bibnamefont {Zoller}},\ }\bibfield  {title} {\bibinfo {title} {Dipole
  {Blockade} and {Quantum} {Information} {Processing} in {Mesoscopic} {Atomic}
  {Ensembles}},\ }\href@noop {} {\bibfield  {journal} {\bibinfo  {journal}
  {Physical Review Letters}\ }\textbf {\bibinfo {volume} {87}},\ \bibinfo
  {pages} {037901} (\bibinfo {year} {2001})}\BibitemShut {NoStop}%
\bibitem [{\citenamefont {Saffman}\ \emph {et~al.}(2010)\citenamefont
  {Saffman}, \citenamefont {Walker},\ and\ \citenamefont
  {Mølmer}}]{Saffman2010c}%
  \BibitemOpen
  \bibfield  {author} {\bibinfo {author} {\bibfnamefont {M.}~\bibnamefont
  {Saffman}}, \bibinfo {author} {\bibfnamefont {T.~G.}\ \bibnamefont
  {Walker}},\ and\ \bibinfo {author} {\bibfnamefont {K.}~\bibnamefont
  {Mølmer}},\ }\bibfield  {title} {\bibinfo {title} {Quantum information with
  {Rydberg} atoms},\ }\href@noop {} {\bibfield  {journal} {\bibinfo  {journal}
  {Reviews of Modern Physics}\ }\textbf {\bibinfo {volume} {82}},\ \bibinfo
  {pages} {2313} (\bibinfo {year} {2010})}\BibitemShut {NoStop}%
\bibitem [{\citenamefont {Peyronel}\ \emph {et~al.}(2012)\citenamefont
  {Peyronel}, \citenamefont {Firstenberg}, \citenamefont {Liang}, \citenamefont
  {Hofferberth}, \citenamefont {Gorshkov}, \citenamefont {Pohl}, \citenamefont
  {Lukin},\ and\ \citenamefont {Vuletić}}]{Peyronel2012a}%
  \BibitemOpen
  \bibfield  {author} {\bibinfo {author} {\bibfnamefont {T.}~\bibnamefont
  {Peyronel}}, \bibinfo {author} {\bibfnamefont {O.}~\bibnamefont
  {Firstenberg}}, \bibinfo {author} {\bibfnamefont {Q.-Y.}\ \bibnamefont
  {Liang}}, \bibinfo {author} {\bibfnamefont {S.}~\bibnamefont {Hofferberth}},
  \bibinfo {author} {\bibfnamefont {A.~V.}\ \bibnamefont {Gorshkov}}, \bibinfo
  {author} {\bibfnamefont {T.}~\bibnamefont {Pohl}}, \bibinfo {author}
  {\bibfnamefont {M.~D.}\ \bibnamefont {Lukin}},\ and\ \bibinfo {author}
  {\bibfnamefont {V.}~\bibnamefont {Vuletić}},\ }\bibfield  {title} {\bibinfo
  {title} {Quantum nonlinear optics with single photons enabled by strongly
  interacting atoms},\ }\href@noop {} {\bibfield  {journal} {\bibinfo
  {journal} {Nature}\ }\textbf {\bibinfo {volume} {488}},\ \bibinfo {pages}
  {57} (\bibinfo {year} {2012})}\BibitemShut {NoStop}%
\bibitem [{\citenamefont {Dudin}\ \emph {et~al.}(2012)\citenamefont {Dudin},
  \citenamefont {Li}, \citenamefont {Bariani},\ and\ \citenamefont
  {Kuzmich}}]{Dudin2012e}%
  \BibitemOpen
  \bibfield  {author} {\bibinfo {author} {\bibfnamefont {Y.~O.}\ \bibnamefont
  {Dudin}}, \bibinfo {author} {\bibfnamefont {L.}~\bibnamefont {Li}}, \bibinfo
  {author} {\bibfnamefont {F.}~\bibnamefont {Bariani}},\ and\ \bibinfo {author}
  {\bibfnamefont {A.}~\bibnamefont {Kuzmich}},\ }\bibfield  {title} {\bibinfo
  {title} {Observation of coherent many-body {Rabi} oscillations},\ }\href@noop
  {} {\bibfield  {journal} {\bibinfo  {journal} {Nature Physics}\ }\textbf
  {\bibinfo {volume} {8}},\ \bibinfo {pages} {790} (\bibinfo {year}
  {2012})}\BibitemShut {NoStop}%
\bibitem [{\citenamefont {Dudin}\ and\ \citenamefont
  {Kuzmich}(2012)}]{Dudin2012sc}%
  \BibitemOpen
  \bibfield  {author} {\bibinfo {author} {\bibfnamefont {Y.~O.}\ \bibnamefont
  {Dudin}}\ and\ \bibinfo {author} {\bibfnamefont {A.}~\bibnamefont
  {Kuzmich}},\ }\bibfield  {title} {\bibinfo {title} {{Strongly Interacting
  Rydberg Excitations of a Cold Atomic Gas}},\ }\href@noop {} {\bibfield
  {journal} {\bibinfo  {journal} {Science}\ }\textbf {\bibinfo {volume}
  {336}},\ \bibinfo {pages} {887} (\bibinfo {year} {2012})}\BibitemShut
  {NoStop}%
\bibitem [{\citenamefont {Firstenberg}\ \emph {et~al.}(2013)\citenamefont
  {Firstenberg}, \citenamefont {Peyronel}, \citenamefont {Liang}, \citenamefont
  {Gorshkov}, \citenamefont {Lukin},\ and\ \citenamefont
  {Vuletić}}]{Firstenberg2013a}%
  \BibitemOpen
  \bibfield  {author} {\bibinfo {author} {\bibfnamefont {O.}~\bibnamefont
  {Firstenberg}}, \bibinfo {author} {\bibfnamefont {T.}~\bibnamefont
  {Peyronel}}, \bibinfo {author} {\bibfnamefont {Q.-Y.}\ \bibnamefont {Liang}},
  \bibinfo {author} {\bibfnamefont {A.~V.}\ \bibnamefont {Gorshkov}}, \bibinfo
  {author} {\bibfnamefont {M.~D.}\ \bibnamefont {Lukin}},\ and\ \bibinfo
  {author} {\bibfnamefont {V.}~\bibnamefont {Vuletić}},\ }\bibfield  {title}
  {\bibinfo {title} {Attractive photons in a quantum nonlinear medium},\
  }\href@noop {} {\bibfield  {journal} {\bibinfo  {journal} {Nature}\ }\textbf
  {\bibinfo {volume} {502}},\ \bibinfo {pages} {71} (\bibinfo {year}
  {2013})}\BibitemShut {NoStop}%
\bibitem [{\citenamefont {Li}\ \emph {et~al.}(2013)\citenamefont {Li},
  \citenamefont {Dudin},\ and\ \citenamefont {Kuzmich}}]{Li2013}%
  \BibitemOpen
  \bibfield  {author} {\bibinfo {author} {\bibfnamefont {L.}~\bibnamefont
  {Li}}, \bibinfo {author} {\bibfnamefont {Y.~O.}\ \bibnamefont {Dudin}},\ and\
  \bibinfo {author} {\bibfnamefont {A.}~\bibnamefont {Kuzmich}},\ }\bibfield
  {title} {\bibinfo {title} {{Entanglement between light and an optical atomic
  excitation}},\ }\href@noop {} {\bibfield  {journal} {\bibinfo  {journal}
  {Nature}\ }\textbf {\bibinfo {volume} {498}},\ \bibinfo {pages} {466}
  (\bibinfo {year} {2013})}\BibitemShut {NoStop}%
\bibitem [{\citenamefont {Tiarks}\ \emph {et~al.}(2014)\citenamefont {Tiarks},
  \citenamefont {Baur}, \citenamefont {Schneider}, \citenamefont {Dürr},\ and\
  \citenamefont {Rempe}}]{tiarks_single-photon_2014}%
  \BibitemOpen
  \bibfield  {author} {\bibinfo {author} {\bibfnamefont {D.}~\bibnamefont
  {Tiarks}}, \bibinfo {author} {\bibfnamefont {S.}~\bibnamefont {Baur}},
  \bibinfo {author} {\bibfnamefont {K.}~\bibnamefont {Schneider}}, \bibinfo
  {author} {\bibfnamefont {S.}~\bibnamefont {Dürr}},\ and\ \bibinfo {author}
  {\bibfnamefont {G.}~\bibnamefont {Rempe}},\ }\bibfield  {title} {\bibinfo
  {title} {Single-{Photon} {Transistor} {Using} a {Forster} {Resonance}},\
  }\href@noop {} {\bibfield  {journal} {\bibinfo  {journal} {Physical Review
  Letters}\ }\textbf {\bibinfo {volume} {113}},\ \bibinfo {pages} {053602}
  (\bibinfo {year} {2014})}\BibitemShut {NoStop}%
\bibitem [{\citenamefont {Gorniaczyk}\ \emph {et~al.}(2014)\citenamefont
  {Gorniaczyk}, \citenamefont {Tresp}, \citenamefont {Schmidt}, \citenamefont
  {Fedder},\ and\ \citenamefont {Hofferberth}}]{gorniaczyk_single-photon_2014}%
  \BibitemOpen
  \bibfield  {author} {\bibinfo {author} {\bibfnamefont {H.}~\bibnamefont
  {Gorniaczyk}}, \bibinfo {author} {\bibfnamefont {C.}~\bibnamefont {Tresp}},
  \bibinfo {author} {\bibfnamefont {J.}~\bibnamefont {Schmidt}}, \bibinfo
  {author} {\bibfnamefont {H.}~\bibnamefont {Fedder}},\ and\ \bibinfo {author}
  {\bibfnamefont {S.}~\bibnamefont {Hofferberth}},\ }\bibfield  {title}
  {\bibinfo {title} {Single-{Photon} {Transistor} {Mediated} by {Interstate}
  {Rydberg} {Interactions}},\ }\href@noop {} {\bibfield  {journal} {\bibinfo
  {journal} {Physical Review Letters}\ }\textbf {\bibinfo {volume} {113}},\
  \bibinfo {pages} {053601} (\bibinfo {year} {2014})}\BibitemShut {NoStop}%
\bibitem [{\citenamefont {Ebert}\ \emph {et~al.}(2014)\citenamefont {Ebert},
  \citenamefont {Gill}, \citenamefont {Gibbons}, \citenamefont {Zhang},
  \citenamefont {Saffman},\ and\ \citenamefont {Walker}}]{Ebert2014}%
  \BibitemOpen
  \bibfield  {author} {\bibinfo {author} {\bibfnamefont {M.}~\bibnamefont
  {Ebert}}, \bibinfo {author} {\bibfnamefont {A.}~\bibnamefont {Gill}},
  \bibinfo {author} {\bibfnamefont {M.}~\bibnamefont {Gibbons}}, \bibinfo
  {author} {\bibfnamefont {X.}~\bibnamefont {Zhang}}, \bibinfo {author}
  {\bibfnamefont {M.}~\bibnamefont {Saffman}},\ and\ \bibinfo {author}
  {\bibfnamefont {T.~G.}\ \bibnamefont {Walker}},\ }\bibfield  {title}
  {\bibinfo {title} {Atomic {Fock} {State} {Preparation} {Using} {Rydberg}
  {Blockade}},\ }\href@noop {} {\bibfield  {journal} {\bibinfo  {journal}
  {Physical Review Letters}\ }\textbf {\bibinfo {volume} {112}},\ \bibinfo
  {pages} {043602} (\bibinfo {year} {2014})}\BibitemShut {NoStop}%
\bibitem [{\citenamefont {Ebert}\ \emph {et~al.}(2015)\citenamefont {Ebert},
  \citenamefont {Kwon}, \citenamefont {Walker},\ and\ \citenamefont
  {Saffman}}]{ebert_coherence_2015}%
  \BibitemOpen
  \bibfield  {author} {\bibinfo {author} {\bibfnamefont {M.}~\bibnamefont
  {Ebert}}, \bibinfo {author} {\bibfnamefont {M.}~\bibnamefont {Kwon}},
  \bibinfo {author} {\bibfnamefont {T.}~\bibnamefont {Walker}},\ and\ \bibinfo
  {author} {\bibfnamefont {M.}~\bibnamefont {Saffman}},\ }\bibfield  {title}
  {\bibinfo {title} {Coherence and {Rydberg} {Blockade} of {Atomic} {Ensemble}
  {Qubits}},\ }\href@noop {} {\bibfield  {journal} {\bibinfo  {journal}
  {Physical Review Letters}\ }\textbf {\bibinfo {volume} {115}},\ \bibinfo
  {pages} {093601} (\bibinfo {year} {2015})}\BibitemShut {NoStop}%
\bibitem [{\citenamefont {Li}\ \emph {et~al.}(2016)\citenamefont {Li},
  \citenamefont {Zhou}, \citenamefont {Jing}, \citenamefont {Wang},
  \citenamefont {Yang}, \citenamefont {Jiang}, \citenamefont {Mølmer},
  \citenamefont {Bao},\ and\ \citenamefont {Pan}}]{Li2016}%
  \BibitemOpen
  \bibfield  {author} {\bibinfo {author} {\bibfnamefont {J.}~\bibnamefont
  {Li}}, \bibinfo {author} {\bibfnamefont {M.-T.}\ \bibnamefont {Zhou}},
  \bibinfo {author} {\bibfnamefont {B.}~\bibnamefont {Jing}}, \bibinfo {author}
  {\bibfnamefont {X.-J.}\ \bibnamefont {Wang}}, \bibinfo {author}
  {\bibfnamefont {S.-J.}\ \bibnamefont {Yang}}, \bibinfo {author}
  {\bibfnamefont {X.}~\bibnamefont {Jiang}}, \bibinfo {author} {\bibfnamefont
  {K.}~\bibnamefont {Mølmer}}, \bibinfo {author} {\bibfnamefont {X.-H.}\
  \bibnamefont {Bao}},\ and\ \bibinfo {author} {\bibfnamefont {J.-W.}\
  \bibnamefont {Pan}},\ }\bibfield  {title} {\bibinfo {title}
  {Hong-{Ou}-{Mandel} {Interference} between {Two} {Deterministic} {Collective}
  {Excitations} in an {Atomic} {Ensemble}},\ }\href@noop {} {\bibfield
  {journal} {\bibinfo  {journal} {Physical Review Letters}\ }\textbf {\bibinfo
  {volume} {117}},\ \bibinfo {pages} {180501} (\bibinfo {year}
  {2016})}\BibitemShut {NoStop}%
\bibitem [{\citenamefont {Gorniaczyk}\ \emph {et~al.}(2016)\citenamefont
  {Gorniaczyk}, \citenamefont {Tresp}, \citenamefont {Bienias}, \citenamefont
  {Paris-Mandoki}, \citenamefont {Li}, \citenamefont {Mirgorodskiy},
  \citenamefont {Büchler}, \citenamefont {Lesanovsky},\ and\ \citenamefont
  {Hofferberth}}]{gorniaczyk_enhancement_2016}%
  \BibitemOpen
  \bibfield  {author} {\bibinfo {author} {\bibfnamefont {H.}~\bibnamefont
  {Gorniaczyk}}, \bibinfo {author} {\bibfnamefont {C.}~\bibnamefont {Tresp}},
  \bibinfo {author} {\bibfnamefont {P.}~\bibnamefont {Bienias}}, \bibinfo
  {author} {\bibfnamefont {A.}~\bibnamefont {Paris-Mandoki}}, \bibinfo {author}
  {\bibfnamefont {W.}~\bibnamefont {Li}}, \bibinfo {author} {\bibfnamefont
  {I.}~\bibnamefont {Mirgorodskiy}}, \bibinfo {author} {\bibfnamefont {H.~P.}\
  \bibnamefont {Büchler}}, \bibinfo {author} {\bibfnamefont {I.}~\bibnamefont
  {Lesanovsky}},\ and\ \bibinfo {author} {\bibfnamefont {S.}~\bibnamefont
  {Hofferberth}},\ }\bibfield  {title} {\bibinfo {title} {Enhancement of
  {Rydberg}-mediated single-photon nonlinearities by electrically tuned
  {Förster} resonances},\ }\href@noop {} {\bibfield  {journal} {\bibinfo
  {journal} {Nature Communications}\ }\textbf {\bibinfo {volume} {7}},\
  \bibinfo {pages} {12480} (\bibinfo {year} {2016})}\BibitemShut {NoStop}%
\bibitem [{\citenamefont {Thompson}\ \emph {et~al.}(2017)\citenamefont
  {Thompson}, \citenamefont {Nicholson}, \citenamefont {Liang}, \citenamefont
  {Cantu}, \citenamefont {Venkatramani}, \citenamefont {Choi}, \citenamefont
  {Fedorov}, \citenamefont {Viscor}, \citenamefont {Pohl}, \citenamefont
  {Lukin},\ and\ \citenamefont {Vuletić}}]{thompson_symmetry-protected_2017}%
  \BibitemOpen
  \bibfield  {author} {\bibinfo {author} {\bibfnamefont {J.~D.}\ \bibnamefont
  {Thompson}}, \bibinfo {author} {\bibfnamefont {T.~L.}\ \bibnamefont
  {Nicholson}}, \bibinfo {author} {\bibfnamefont {Q.-Y.}\ \bibnamefont
  {Liang}}, \bibinfo {author} {\bibfnamefont {S.~H.}\ \bibnamefont {Cantu}},
  \bibinfo {author} {\bibfnamefont {A.~V.}\ \bibnamefont {Venkatramani}},
  \bibinfo {author} {\bibfnamefont {S.}~\bibnamefont {Choi}}, \bibinfo {author}
  {\bibfnamefont {I.~A.}\ \bibnamefont {Fedorov}}, \bibinfo {author}
  {\bibfnamefont {D.}~\bibnamefont {Viscor}}, \bibinfo {author} {\bibfnamefont
  {T.}~\bibnamefont {Pohl}}, \bibinfo {author} {\bibfnamefont {M.~D.}\
  \bibnamefont {Lukin}},\ and\ \bibinfo {author} {\bibfnamefont
  {V.}~\bibnamefont {Vuletić}},\ }\bibfield  {title} {\bibinfo {title}
  {Symmetry-protected collisions between strongly interacting photons},\
  }\href@noop {} {\bibfield  {journal} {\bibinfo  {journal} {Nature}\ }\textbf
  {\bibinfo {volume} {542}},\ \bibinfo {pages} {206} (\bibinfo {year}
  {2017})}\BibitemShut {NoStop}%
\bibitem [{\citenamefont {Busche}\ \emph {et~al.}(2017)\citenamefont {Busche},
  \citenamefont {Huillery}, \citenamefont {Ball}, \citenamefont {Ilieva},
  \citenamefont {Jones},\ and\ \citenamefont
  {Adams}}]{busche_contactless_2017}%
  \BibitemOpen
  \bibfield  {author} {\bibinfo {author} {\bibfnamefont {H.}~\bibnamefont
  {Busche}}, \bibinfo {author} {\bibfnamefont {P.}~\bibnamefont {Huillery}},
  \bibinfo {author} {\bibfnamefont {S.~W.}\ \bibnamefont {Ball}}, \bibinfo
  {author} {\bibfnamefont {T.}~\bibnamefont {Ilieva}}, \bibinfo {author}
  {\bibfnamefont {M.~P.~A.}\ \bibnamefont {Jones}},\ and\ \bibinfo {author}
  {\bibfnamefont {C.~S.}\ \bibnamefont {Adams}},\ }\bibfield  {title} {\bibinfo
  {title} {Contactless nonlinear optics mediated by long-range {Rydberg}
  interactions},\ }\href@noop {} {\bibfield  {journal} {\bibinfo  {journal}
  {Nature Physics}\ }\textbf {\bibinfo {volume} {13}},\ \bibinfo {pages} {655}
  (\bibinfo {year} {2017})}\BibitemShut {NoStop}%
\bibitem [{\citenamefont {Tiarks}\ \emph {et~al.}(2019)\citenamefont {Tiarks},
  \citenamefont {Schmidt-Eberle}, \citenamefont {Stolz}, \citenamefont
  {Rempe},\ and\ \citenamefont {Dürr}}]{tiarks_photonphoton_2019}%
  \BibitemOpen
  \bibfield  {author} {\bibinfo {author} {\bibfnamefont {D.}~\bibnamefont
  {Tiarks}}, \bibinfo {author} {\bibfnamefont {S.}~\bibnamefont
  {Schmidt-Eberle}}, \bibinfo {author} {\bibfnamefont {T.}~\bibnamefont
  {Stolz}}, \bibinfo {author} {\bibfnamefont {G.}~\bibnamefont {Rempe}},\ and\
  \bibinfo {author} {\bibfnamefont {S.}~\bibnamefont {Dürr}},\ }\bibfield
  {title} {\bibinfo {title} {A photon–photon quantum gate based on {Rydberg}
  interactions},\ }\href@noop {} {\bibfield  {journal} {\bibinfo  {journal}
  {Nature Physics}\ }\textbf {\bibinfo {volume} {15}},\ \bibinfo {pages} {124}
  (\bibinfo {year} {2019})}\BibitemShut {NoStop}%
\bibitem [{\citenamefont {Li}\ \emph {et~al.}(2019)\citenamefont {Li},
  \citenamefont {Zhou}, \citenamefont {Yang}, \citenamefont {Sun},
  \citenamefont {Liu}, \citenamefont {Bao},\ and\ \citenamefont
  {Pan}}]{Li2019}%
  \BibitemOpen
  \bibfield  {author} {\bibinfo {author} {\bibfnamefont {J.}~\bibnamefont
  {Li}}, \bibinfo {author} {\bibfnamefont {M.-T.}\ \bibnamefont {Zhou}},
  \bibinfo {author} {\bibfnamefont {C.-W.}\ \bibnamefont {Yang}}, \bibinfo
  {author} {\bibfnamefont {P.-F.}\ \bibnamefont {Sun}}, \bibinfo {author}
  {\bibfnamefont {J.-L.}\ \bibnamefont {Liu}}, \bibinfo {author} {\bibfnamefont
  {X.-H.}\ \bibnamefont {Bao}},\ and\ \bibinfo {author} {\bibfnamefont {J.-W.}\
  \bibnamefont {Pan}},\ }\bibfield  {title} {\bibinfo {title}
  {{Semideterministic Entanglement between a Single Photon and an Atomic
  Ensemble}},\ }\href@noop {} {\bibfield  {journal} {\bibinfo  {journal}
  {Physical Review Letters}\ }\textbf {\bibinfo {volume} {123}},\ \bibinfo
  {pages} {140504} (\bibinfo {year} {2019})}\BibitemShut {NoStop}%
\bibitem [{\citenamefont {Cantu}\ \emph {et~al.}(2020)\citenamefont {Cantu},
  \citenamefont {Venkatramani}, \citenamefont {Xu}, \citenamefont {Zhou},
  \citenamefont {Jelenković}, \citenamefont {Lukin},\ and\ \citenamefont
  {Vuletić}}]{cantu_repulsive_2020}%
  \BibitemOpen
  \bibfield  {author} {\bibinfo {author} {\bibfnamefont {S.~H.}\ \bibnamefont
  {Cantu}}, \bibinfo {author} {\bibfnamefont {A.~V.}\ \bibnamefont
  {Venkatramani}}, \bibinfo {author} {\bibfnamefont {W.}~\bibnamefont {Xu}},
  \bibinfo {author} {\bibfnamefont {L.}~\bibnamefont {Zhou}}, \bibinfo {author}
  {\bibfnamefont {B.}~\bibnamefont {Jelenković}}, \bibinfo {author}
  {\bibfnamefont {M.~D.}\ \bibnamefont {Lukin}},\ and\ \bibinfo {author}
  {\bibfnamefont {V.}~\bibnamefont {Vuletić}},\ }\bibfield  {title} {\bibinfo
  {title} {Repulsive photons in a quantum nonlinear medium},\ }\href@noop {}
  {\bibfield  {journal} {\bibinfo  {journal} {Nature Physics}\ }\textbf
  {\bibinfo {volume} {16}},\ \bibinfo {pages} {921} (\bibinfo {year}
  {2020})}\BibitemShut {NoStop}%
\bibitem [{\citenamefont {Yuan}\ \emph {et~al.}(2008)\citenamefont {Yuan},
  \citenamefont {Chen}, \citenamefont {Zhao}, \citenamefont {Chen},
  \citenamefont {Schmiedmayer},\ and\ \citenamefont
  {Pan}}]{yuan2008experimental}%
  \BibitemOpen
  \bibfield  {author} {\bibinfo {author} {\bibfnamefont {Z.-S.}\ \bibnamefont
  {Yuan}}, \bibinfo {author} {\bibfnamefont {Y.-A.}\ \bibnamefont {Chen}},
  \bibinfo {author} {\bibfnamefont {B.}~\bibnamefont {Zhao}}, \bibinfo {author}
  {\bibfnamefont {S.}~\bibnamefont {Chen}}, \bibinfo {author} {\bibfnamefont
  {J.}~\bibnamefont {Schmiedmayer}},\ and\ \bibinfo {author} {\bibfnamefont
  {J.-W.}\ \bibnamefont {Pan}},\ }\bibfield  {title} {\bibinfo {title}
  {Experimental demonstration of a bdcz quantum repeater node},\ }\href@noop {}
  {\bibfield  {journal} {\bibinfo  {journal} {Nature}\ }\textbf {\bibinfo
  {volume} {454}},\ \bibinfo {pages} {1098} (\bibinfo {year}
  {2008})}\BibitemShut {NoStop}%
\bibitem [{\citenamefont {Pan}\ \emph {et~al.}(1998)\citenamefont {Pan},
  \citenamefont {Bouwmeester}, \citenamefont {Weinfurter},\ and\ \citenamefont
  {Zeilinger}}]{Pan1998}%
  \BibitemOpen
  \bibfield  {author} {\bibinfo {author} {\bibfnamefont {J.-W.}\ \bibnamefont
  {Pan}}, \bibinfo {author} {\bibfnamefont {D.}~\bibnamefont {Bouwmeester}},
  \bibinfo {author} {\bibfnamefont {H.}~\bibnamefont {Weinfurter}},\ and\
  \bibinfo {author} {\bibfnamefont {A.}~\bibnamefont {Zeilinger}},\ }\bibfield
  {title} {\bibinfo {title} {{Experimental Entanglement Swapping: Entangling
  Photons That Never Interacted}},\ }\href@noop {} {\bibfield  {journal}
  {\bibinfo  {journal} {Physical Review Letters}\ }\textbf {\bibinfo {volume}
  {80}},\ \bibinfo {pages} {3891} (\bibinfo {year} {1998})}\BibitemShut
  {NoStop}%
\bibitem [{\citenamefont {Saffman}\ and\ \citenamefont
  {Walker}(2002)}]{Saffman2002}%
  \BibitemOpen
  \bibfield  {author} {\bibinfo {author} {\bibfnamefont {M.}~\bibnamefont
  {Saffman}}\ and\ \bibinfo {author} {\bibfnamefont {T.~G.}\ \bibnamefont
  {Walker}},\ }\bibfield  {title} {\bibinfo {title} {Creating single-atom and
  single-photon sources from entangled atomic ensembles},\ }\href@noop {}
  {\bibfield  {journal} {\bibinfo  {journal} {Physical Review A}\ }\textbf
  {\bibinfo {volume} {66}},\ \bibinfo {pages} {065403} (\bibinfo {year}
  {2002})}\BibitemShut {NoStop}%
\bibitem [{\citenamefont {Zhou}\ \emph {et~al.}(2020)\citenamefont {Zhou},
  \citenamefont {Liu}, \citenamefont {Sun}, \citenamefont {An}, \citenamefont
  {Li}, \citenamefont {Bao},\ and\ \citenamefont {Pan}}]{Zhou2020}%
  \BibitemOpen
  \bibfield  {author} {\bibinfo {author} {\bibfnamefont {M.-T.}\ \bibnamefont
  {Zhou}}, \bibinfo {author} {\bibfnamefont {J.-L.}\ \bibnamefont {Liu}},
  \bibinfo {author} {\bibfnamefont {P.-F.}\ \bibnamefont {Sun}}, \bibinfo
  {author} {\bibfnamefont {Z.-Y.}\ \bibnamefont {An}}, \bibinfo {author}
  {\bibfnamefont {J.}~\bibnamefont {Li}}, \bibinfo {author} {\bibfnamefont
  {X.-H.}\ \bibnamefont {Bao}},\ and\ \bibinfo {author} {\bibfnamefont {J.-W.}\
  \bibnamefont {Pan}},\ }\bibfield  {title} {\bibinfo {title} {Experimental
  creation of single rydberg excitations via adiabatic passage},\ }\href@noop
  {} {\bibfield  {journal} {\bibinfo  {journal} {Physical Review A}\ }\textbf
  {\bibinfo {volume} {102}},\ \bibinfo {pages} {013706} (\bibinfo {year}
  {2020})}\BibitemShut {NoStop}%
\bibitem [{\citenamefont {Ketterle}\ \emph {et~al.}(1993)\citenamefont
  {Ketterle}, \citenamefont {Davis}, \citenamefont {Joffe}, \citenamefont
  {Martin},\ and\ \citenamefont {Pritchard}}]{Ketterle1993}%
  \BibitemOpen
  \bibfield  {author} {\bibinfo {author} {\bibfnamefont {W.}~\bibnamefont
  {Ketterle}}, \bibinfo {author} {\bibfnamefont {K.~B.}\ \bibnamefont {Davis}},
  \bibinfo {author} {\bibfnamefont {M.~A.}\ \bibnamefont {Joffe}}, \bibinfo
  {author} {\bibfnamefont {A.}~\bibnamefont {Martin}},\ and\ \bibinfo {author}
  {\bibfnamefont {D.~E.}\ \bibnamefont {Pritchard}},\ }\bibfield  {title}
  {\bibinfo {title} {High densities of cold atoms in a dark spontaneous-force
  optical trap},\ }\href@noop {} {\bibfield  {journal} {\bibinfo  {journal}
  {Physical Review Letters}\ }\textbf {\bibinfo {volume} {70}},\ \bibinfo
  {pages} {2253} (\bibinfo {year} {1993})}\BibitemShut {NoStop}%
\bibitem [{\citenamefont {Townsend}\ \emph {et~al.}(1996)\citenamefont
  {Townsend}, \citenamefont {Edwards}, \citenamefont {Zetie}, \citenamefont
  {Cooper}, \citenamefont {Rink},\ and\ \citenamefont {Foot}}]{Townsend1996}%
  \BibitemOpen
  \bibfield  {author} {\bibinfo {author} {\bibfnamefont {C.}~\bibnamefont
  {Townsend}}, \bibinfo {author} {\bibfnamefont {N.}~\bibnamefont {Edwards}},
  \bibinfo {author} {\bibfnamefont {K.}~\bibnamefont {Zetie}}, \bibinfo
  {author} {\bibfnamefont {C.}~\bibnamefont {Cooper}}, \bibinfo {author}
  {\bibfnamefont {J.}~\bibnamefont {Rink}},\ and\ \bibinfo {author}
  {\bibfnamefont {C.}~\bibnamefont {Foot}},\ }\bibfield  {title} {\bibinfo
  {title} {High-density trapping of cesium atoms in a dark magneto-optical
  trap},\ }\href@noop {} {\bibfield  {journal} {\bibinfo  {journal} {Physical
  Review A}\ }\textbf {\bibinfo {volume} {53}},\ \bibinfo {pages} {1702}
  (\bibinfo {year} {1996})}\BibitemShut {NoStop}%
\bibitem [{\citenamefont {Gühne}\ and\ \citenamefont
  {Tóth}(2009)}]{Guhne2009}%
  \BibitemOpen
  \bibfield  {author} {\bibinfo {author} {\bibfnamefont {O.}~\bibnamefont
  {Gühne}}\ and\ \bibinfo {author} {\bibfnamefont {G.}~\bibnamefont {Tóth}},\
  }\bibfield  {title} {\bibinfo {title} {Entanglement detection},\ }\href@noop
  {} {\bibfield  {journal} {\bibinfo  {journal} {Physics Reports}\ }\textbf
  {\bibinfo {volume} {474}},\ \bibinfo {pages} {1} (\bibinfo {year}
  {2009})}\BibitemShut {NoStop}%
\bibitem [{\citenamefont {Bell}(1966)}]{bell_problem_1966}%
  \BibitemOpen
  \bibfield  {author} {\bibinfo {author} {\bibfnamefont {J.~S.}\ \bibnamefont
  {Bell}},\ }\bibfield  {title} {\bibinfo {title} {On the {Problem} of {Hidden}
  {Variables} in {Quantum} {Mechanics}},\ }\href@noop {} {\bibfield  {journal}
  {\bibinfo  {journal} {Reviews of Modern Physics}\ }\textbf {\bibinfo {volume}
  {38}},\ \bibinfo {pages} {447} (\bibinfo {year} {1966})}\BibitemShut
  {NoStop}%
\bibitem [{\citenamefont {Clauser}\ \emph {et~al.}(1969)\citenamefont
  {Clauser}, \citenamefont {Horne}, \citenamefont {Shimony},\ and\
  \citenamefont {Holt}}]{clauser_proposed_1969}%
  \BibitemOpen
  \bibfield  {author} {\bibinfo {author} {\bibfnamefont {J.~F.}\ \bibnamefont
  {Clauser}}, \bibinfo {author} {\bibfnamefont {M.~A.}\ \bibnamefont {Horne}},
  \bibinfo {author} {\bibfnamefont {A.}~\bibnamefont {Shimony}},\ and\ \bibinfo
  {author} {\bibfnamefont {R.~A.}\ \bibnamefont {Holt}},\ }\bibfield  {title}
  {\bibinfo {title} {Proposed {Experiment} to {Test} {Local}
  {Hidden}-{Variable} {Theories}},\ }\href@noop {} {\bibfield  {journal}
  {\bibinfo  {journal} {Physical Review Letters}\ }\textbf {\bibinfo {volume}
  {23}},\ \bibinfo {pages} {880} (\bibinfo {year} {1969})}\BibitemShut
  {NoStop}%
\end{thebibliography}%

\end{document}